%% file: Main.tex
\RequirePackage{fix-cm}

\documentclass[smallextended]{svjour3}

\pdfoutput=1

\usepackage{lineno} 
\usepackage{tcolorbox}
\usepackage{latexsym}
\usepackage[center]{caption}
\usepackage{multirow}
\usepackage{graphicx}
\usepackage{url}
\usepackage{hyperref}
\usepackage[utf8]{inputenc}
\usepackage{amssymb}
\usepackage{color}
\usepackage{fancybox}
\usepackage{subcaption}
\usepackage{url}
\usepackage{xspace}
\usepackage{amssymb}
\usepackage{multirow}
\usepackage{tabularx}
\graphicspath{{images/}}
\usepackage{url}
\usepackage{tablefootnote}
\usepackage{balance}

\usepackage{xcolor}

\modulolinenumbers[5]

\usepackage{amssymb}
\makeatletter

\newcommand{\blue}[1]{#1}

\newcolumntype{C}{>{\centering\arraybackslash}p{12em}}

\newtcolorbox{mybox}[1]{fonttitle=\bfseries,title=#1}

\newcommand{\al}{\textit{et al.~}}

\newcommand{\eg}{{\textit{e.g.,}}}
\newcommand{\ie}{{\textit{i.e.,}}}
\newcommand{\ea}{{et al.}}
\newcommand{\approach}{{Swarm Debugging~\xspace}}
\newcommand{\acron}{{SD~\xspace}}

\newcommand{\resbox}[1]{\vspace{0.8em}\begin{center}%    
    \noindent\thicklines\setlength{\fboxsep}{8pt}%    
    \cornersize{0.2}\Ovalbox{\begin{minipage}{3.0in}%    
    \textit{#1}\end{minipage}}\end{center}\vspace{0.4em}}

\def\RQOne{Is there a correlation between the time of the first breakpoint and a debugging task’s elapsed time?}
\def\RQTwo{What is the effort in time for setting the first breakpoint in relation to the debugging task’s elapsed time?}
\def\RQThree{Are there consistent, common trends with respect to the types of statements on which developers set breakpoints?}
\def\RQFour{Are there consistent, common trends with respect to the lines, methods, or classes on which developers set breakpoints?}
\def\RQFive{Is Swarm Debugging's Global View useful in terms of supporting debugging tasks?}
\def\RQSix{Is Swarm Debugging's Global View useful in terms of sharing debugging tasks?}

\begin{document}

\title{Swarm Debugging: the Collective Intelligence on Interactive Debugging}

\author{Fabio Petrillo$^1$, Yann-Ga\"{e}l Gu\'{e}h\'{e}neuc$^3$, Marcelo Pimenta$^2$, Carla Dal Sasso Freitas$^2$, Foutse Khomh$^4$}
\institute{$^1$Université du Quebéc à Chicoutimi, $^2$Federal University of Rio Grande do Sul, $^3$Concordia University, $^4$Polytechnique Montreal, Canada}

\date{Received: date / Accepted: date}

 \journalname{arXiv Computer Science}
 
\maketitle

\input{01Abstract}


\input{10Introduction}
\input{20Background}
\input{30SwarmDebugging}
\input{40Studies}
\input{60Discussion}
\input{70Threats}
\input{80RelatedWork}
\input{90Conclusion}

\section{Acknowledgment}
This work has been partially supported by the Natural Sciences and Engineering Research Council of Canada (NSERC), the Brazilian research funding agencies CNPq (National Council for Scientific and Technological Development), and CAPES Foundation (Finance Code 001). \blue{We also acknowledge all the participants in our experiments and the insightful comments from the anonymous reviewers.}

\bibliographystyle{spphys}
\bibliography{references}

\input{95Appendix.tex}

\end{document}

%% file: 01Abstract.tex
\abstract{One of the most important tasks in software maintenance is debugging. To start an interactive debugging session, developers \blue{usually} set breakpoints in an integrated development environment and navigate through different paths in their debuggers. \blue{We started our work by asking} what debugging information is useful to share among developers and study two pieces of information: breakpoints (and their locations) and sessions (debugging paths). To answer our question, we introduce the Swarm Debugging concept to frame the sharing of debugging information, the Swarm Debugging Infrastructure (SDI) with which practitioners and researchers can collect and share data about developers' interactive debugging sessions, and the Swarm Debugging Global View (GV) to display debugging paths. Using the SDI, we conducted a large study with professional developers to understand how developers set breakpoints. Using the GV, we also analyzed professional developers in two studies and collected data about their debugging sessions. Our observations and the answers to our research questions  suggest that sharing and visualizing debugging data can support debugging activities.}

%% file: 10Introduction.tex
\section{Introduction}
\label{sec:introduction}

\begin{quote}
\emph{\textbf{Debug.} To detect, locate, and correct faults in a computer program. Techniques include the use of breakpoints, desk checking, dumps, inspection, reversible execution, single-step operations, and traces.}\newline---IEEE Standard Glossary of SE Terminology, 1990 
\end{quote}

Debugging is a common activity during software development, maintenance, and evolution \cite{Tanenbaum1973}. Developers use debugging tools to detect, locate, and correct faults. Debugging tools can be \emph{interactive} or \emph{automated}.

Interactive debugging tools, \emph{a.k.a. debuggers}, such as \textit{sdb} \cite{Katso1979}, \textit{dbx} \cite{Linton1990}, or \textit{gdb} \cite{StallmanR.Pesch2002}, have been used by developers for decades. Modern debuggers are often integrated in interactive environments, \eg{} DDD \cite{P.Wainwright2010} or the debuggers of Eclipse, NetBeans, IntelliJ IDEA, and Visual Studio. \blue{They} allow developers to navigate through the code, look for locations to place breakpoints, and step over/into statements. While stepping, debuggers can traverse method invocations and allow developers to toggle one or more breakpoints and stop/restart executions. Thus, they allow developers to gain knowledge about programs and the causes of faults to fix them.

Automated debugging tools require both successful and failed runs and do not support programs with interactive inputs \cite{Ko2006}. Consequently, they have not been widely adopted in practice. Moreover, automated debugging approaches are often unable to indicate the ``true" locations of faults \cite{Rossler2012}. Other hybrid tools, such as slicing and query languages, may help developers but there is insufficient evidence that \blue{they help} developers during debugging.

Although Integrated Development Environments (IDEs) encourage developers to work collaboratively, exchanging code through Git or assessing code quality with SonarQube, one activity remains solitary: debugging. \blue{Debugging is still an individual activity, during which, a developer explores the source code of the system under development or maintenance using the debugger provided by an IDE. She steps into hundreds of statements and traverses dozens of method invocations painstakingly to gain an understanding of the system.} \blue{Moreover, within} modern interactive debugging tools, such as those included in Eclipse or IntelliJ, a debugging session cannot start if the developer does not set a breakpoint. Consequently, it is mandatory to set at least one breakpoint to \blue{launch} an interactive debugging session. 

\blue{Several studies have shown that developers} spend over two-thirds of their time investigating code and one-third of this time is spent in debugging \cite{LaToza2010,Ko2005,LaToza2006MaintainingMM}. However, \blue{developers do} not reuse the knowledge accumulated during debugging directly. When debugging is over, they loose \blue{track} of the paths that they followed into the code and of the breakpoints that they toggled. Moreover, they cannot share this knowledge with other developers \blue{easily}. If a fault re-appears in the system or if a new fault similar to a previous one is logged, the developer must restart the exploration from the beginning.

\blue{In fact, debugging tools have not changed substantially in the last 30 years: developers’ primary tools for debugging their programs are still breakpoint debuggers and print statements. Indeed, changing the way developers debug their programs is one of the main motivations of our work. We are convinced that a collaborative way of using contextual information of (previous) debugging sessions to support (future) debugging activities is a very interesting approach.}

Ro{\ss}ler \cite{Rossler2012} advocated for the development of a new family of debugging tools that use contextual information. To build context-aware debugging tools, researchers need an understanding of developers' debugging sessions to use this information as context for their debugging. Thus, researchers need tools to collect and share data about developers' debugging sessions.

Maalej \al{} \cite{Maalej2014} observed that capturing contextual information requires the instrumentation of the IDE and continuous observation of the developers' activities within the IDE. Studies by Storey \al{} \cite{Storey2014} showed that the newer generation of developers, who are proficient in social media, are comfortable with sharing such information. Developers are nowadays open, transparent, eager to share their knowledge, and generally willing to allow information about their activities to be collected by the IDEs automatically \cite{Storey2014}. 

\blue{Considering this context,} we introduce the concept of \textbf{Swarm Debugging (SD)} to (1) capture debugging contextual information, (2) share it, and (3) reuse it across debugging sessions and developers. We build the concept of Swarm Debugging \blue{based} on the idea that many developers, performing debugging sessions independently, are in fact building collective knowledge, which can be shared and reused with adequate support. Thus, we are convinced \blue{that} developers need support to collect, store, and share this knowledge, \ie{} information from and about their debugging sessions, including but not limited to breakpoints locations, visited statements, and traversed paths. \blue{To provide} such support, Swarm Debugging includes (i) the Swarm Debugging Infrastructure (SDI), with which practitioners and researchers can collect and share data about developers' interactive debugging sessions, and (ii) the Swarm Debugging Global View (GV) to display debugging paths.

As a consequence of adopting SD, an interesting question emerges: what debugging information is useful to share among developers \blue{to ease} debugging? Debugging provides a lot of information which could be possibly considered useful to improve software comprehension but we are particularly interested \blue{in} two pieces of debugging information: breakpoints (and their locations) and sessions (debugging paths), because these pieces of information \blue{are essential for} the two main activities during debugging: setting breakpoints and stepping in/over/out statements. 

In general, developers initiate an interactive debugging session by setting a breakpoint. Setting a breakpoint is one of the most frequently used features of IDEs \cite{Beller2017}. To \blue{decide} where to set a breakpoint, \blue{developers use} their observations, recall their experiences with similar debugging tasks and formulate hypotheses about their tasks~\cite{Zhang2013}. Tiarks and R\"{o}hms~\cite{Tiarks2013} observed that developers have difficulties in finding locations for setting the breakpoints, suggesting that this is a \blue{demanding} activity and that supporting developers to set appropriate breakpoints could reduce debugging effort.

\blue{We conducted two sets of studies with the aim of understanding how developers set breakpoints and navigate (step) during debugging sessions. In observational studies, we collected and analyzed more than 10 hours of developers' videos in 45 debugging sessions performed by 28 different, independent developers, containing 307 breakpoints on three software systems. These observational studies help us understand how developers use breakpoints (RQ1 to RQ4).}

\blue{We also conducted with 30 professional developers two studies, a qualitative evaluation and a controlled experiment, to assess whether debugging sessions, shared through our Global View visualisation, support developers in their debugging tasks and is useful for sharing debugging tasks among developers (R5 and RQ6). We collected participants' answers in electronic forms and more than 3 hours of debugging sessions on video.}

This paper has the following contributions:
\begin{itemize}
\item We introduce a novel approach for debugging named Swarm Debugging (SD) based on the concept of Swarm Intelligence and Information Foraging Theory.

\item We present an infrastructure, the Swarm Debugging Infrastructure (SDI), to gather, store, and share data about interactive debugging activities to support SD.

\item We provide evidence about the relation between tasks' elapsed time, developers' expertise, breakpoints setting, and debugging patterns.

\item We present a new visualisation technique, Global View (GV), built on shared debugging sessions by developers \blue{to ease} debugging.

\item We provide evidence about the usefulness of sharing debugging session to ease developers’ debugging.
\end{itemize}

This paper extends our previous works \cite{Petrillo2015,PetrilloQRS2016,PetrilloQRS2017} as follows. First, we summarize the main characteristics of the Swarm Debugging approach, providing a theoretical foundation to Swarm Debugging using Swarm Intelligence and Information Foraging Theory. Second, we present the Swarm Debugging Infrastructure (SDI).
Third, we perform an experiment on the debugging behavior of 30 professional developers to evaluate if sharing debugging sessions supports adequately their debugging tasks.

The remainder of this article is organized as follows. Section \ref{Section: Foundations} provides some fundamentals of  debugging and the foundations of SD: the concepts of swarm intelligence and information foraging theory. Section \ref{Section: Our Approach} describes our approach and its implementation, the Swarm Debugging Infrastructure. Section \ref{sec:Evaluation of SD} presents an experiment to assess the benefits that our SD approach can bring to developers, and Section \ref{sec:Using SDI to Understand Debugging Activities} reports two experiments that were conducted using SDI to understand developers debugging habits. Next, Section \ref{Section: Discussions} discusses implications of our results, while Section \ref{sec:threatsValidity} presents threats to the validity of our study. Section \ref{sec:relatedwork} summarizes related work, and finally, Section \ref{sec:conclusion} concludes the paper and outlines future work.

%% file: 20Background.tex
\section{Background}
\label{Section: Foundations}

%\subsection{About Debugging}
%\label{Section: About Debugging}

This section provides background information about the debugging activity and setting breakpoints. In the following, we use \textbf{failures} as unintended behaviours of a program, \ie{} when the program does something that it should not, and \textbf{faults} as the incorrect statements in source code causing failures. The purpose of debugging is to locate and correct faults, hence to fix failures.

\subsection{Debugging and Interactive Debugging}

The IEEE Standard Glossary of Software Engineering Terminology (see the definition at the beginning of Section \ref{sec:introduction}) defines debugging as the act of detecting, locating, and correcting bugs in a computer program. Debugging techniques include the use of breakpoints, desk checking, dumps, inspection, reversible execution, single-step operations, and traces.

Araki \al~\cite{Araki91-DebuggingFramework} describe \textit{debugging} as a process where developers make hypotheses about the root-cause of a problem or defect and verify these hypotheses by examining different parts of the source code of the program.

\textit{Interactive debugging} consists of using a tool, \ie{} \textit{a debugger} to detect, locate, and correct a fault in a program. It is a process also known as \textit{program animation}, \textit{stepping}, or \textit{following execution}~\cite{Zayour2016}. Developers often refer to this process simply as \textit{debugging}, because several IDEs provide debuggers to support \textit{debugging}. However, it must be noted that while \textit{debugging} is the process of finding faults, \textit{interactive debugging} is one particular debugging approach in which developers use interactive tools. 
Expressions such as \textit{interactive debugging}, \textit{stepping} and \textit{debugging} are used interchangeably, and there is not yet a consensus on what is the best name for this process.

\subsection{Breakpoints and Supporting Mechanisms}

Generally, breakpoints allow \blue{pausing} intentionally the execution of a program for debugging purposes, a means of acquiring knowledge about a program during its execution, for example, to examine the call stack and variable values when the control flow reaches the locations of the breakpoints. Thus, a breakpoint indicates the location (line) in the source code of a program where a pause \blue{occurs} during its execution.  

Depending on the programming language, its run-time environment (in particular the capabilities of its virtual machines if any), and the debuggers, different types of breakpoints may be available to developers. These types include static breakpoints~\cite{Chern:2007}, that pause unconditionally the execution of a program, and dynamic breakpoints~\cite{eclipseConditional}, that pause depending on some conditions or threads or numbers of hits.

\begin{figure*}[ht]
    \centering
    \includegraphics[width=0.90\linewidth]{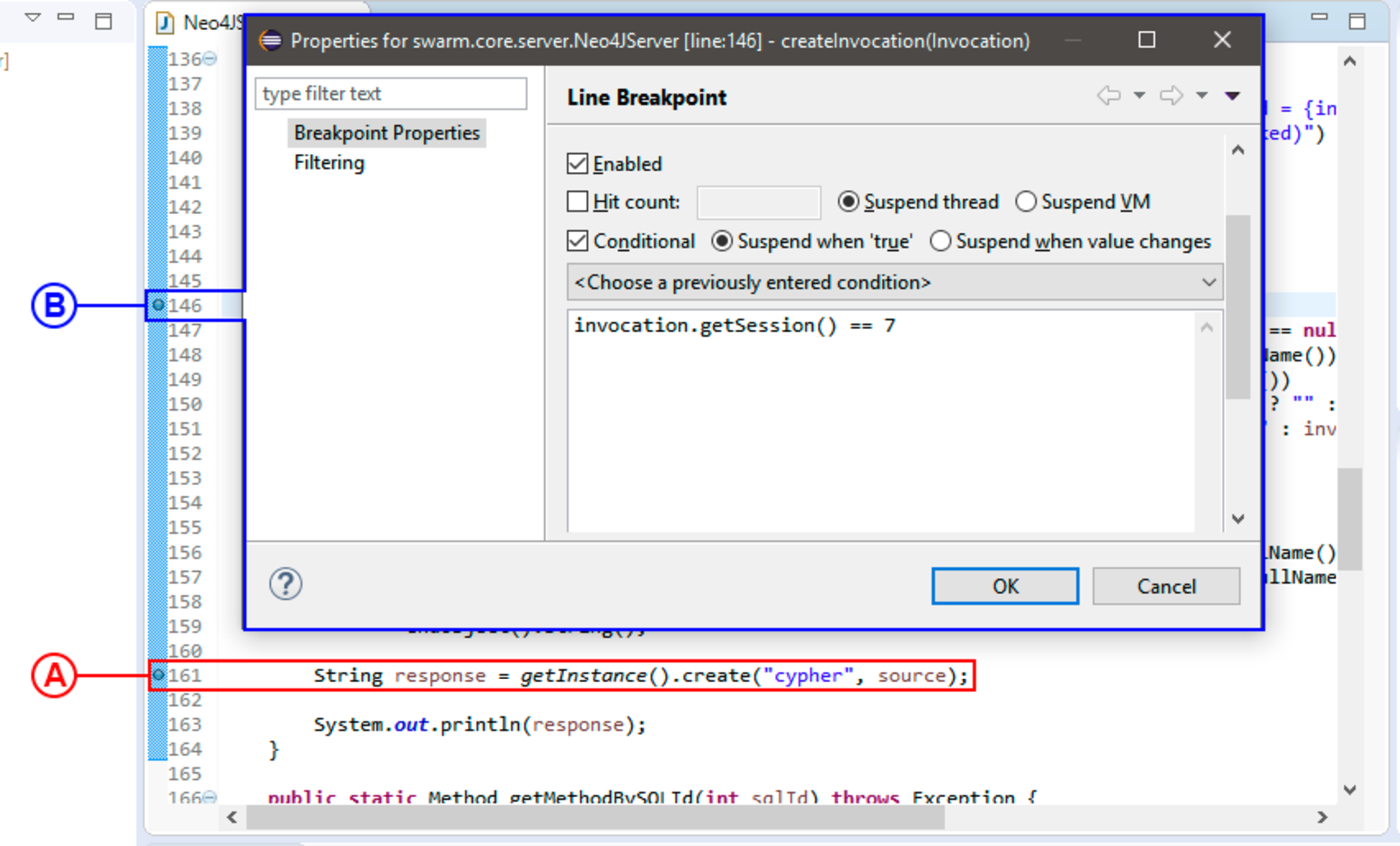}
    \caption{Setting a static breakpoint (A) and a conditional breakpoint (B) using Eclipse IDE}
    \label{fig:eclipse}
\end{figure*}

Other types of breakpoints include watchpoints that pause the execution when a variable being watched is read and--or written. IDEs offer the means to specify the different types of breakpoints depending on the programming languages and their run-time environment. Fig.~\ref{fig:eclipse}-A and~\ref{fig:eclipse}-B show examples of static and dynamic breakpoints in Eclipse. In the rest of this paper, we focus on static breakpoints because they are the most used of all types \cite{Zhang2013}.

There are different mechanisms for setting a breakpoint within the code:

\begin{itemize}
\item \emph{GUI:} Most IDEs or browsers offer a visual way of adding a breakpoint, usually by clicking at the beginning of the line on which to set the breakpoint: \emph{Chrome}\footnote{\tiny{https://developers.google.com/web/tools/chrome-devtools/javascript/add-breakpoints}}, \emph{Visual Studio}\footnote{\tiny{https://msdn.microsoft.com/en-us/library/5557y8b4.aspx}}, \emph{IntelliJ}\footnote{\tiny{https://www.jetbrains.com/help/idea/2016.3/debugger-basics.html}}, and \emph{Xcode}\footnote{\tiny{http://jeffreysambells.com/2014/01/14/using-breakpoints-in-xcode}}.

\item \emph{Command line:} Some programming languages offer debugging tools on the command line, so an IDE is not necessary to debug the code: \emph{JDB}\footnote{\tiny{http://docs.oracle.com/javase/7/docs/technotes/tools/windows/jdb.html}}, \emph{PDB}\footnote{\tiny{https://docs.python.org/2/library/pdb.html}}, and \emph{GDB}\footnote{\tiny{ftp://ftp.gnu.org/old\-gnu/Manuals/gdb\-5.1.1/html\_node/gdb\_37.html}}. 

\item \noindent\emph{Code:} Some programming languages allow using syntactical elements to set breakpoints as they were `annotations' in the code.  This approach often only supports the setting of a breakpoint, and it is necessary to use it in conjunction with the command line or GUI. Some examples are: \emph{Ruby debugger}\footnote{\tiny{https://github.com/cldwalker/debugger}}, \emph{Firefox}\footnote{\tiny{https://developer.mozilla.org/pt-BR/docs/Web/JavaScript/Reference/Statements/debugger}}, and \emph{Chrome}\footnote{\tiny{https://developers.google.com/web/tools/chrome-devtools/javascript/add-breakpoints}}.
\end{itemize}

\noindent There is a set of features in a debugger that allows developers to control the flow of the execution within the breakpoints, \ie{} \textit{Call Stack features}, which enable \emph{continuing} or \emph{stepping}.

A developer can opt for \emph{continuing}, in which case the debugger \blue{resumes} execution until the next breakpoint is reached or the program exits. Conversely, \emph{stepping} allows the developer to run step by step the entire program flow. The definition of a step varies across programming languages and debuggers, but it generally includes invoking a method and executing a statement. While Stepping, a developer can navigate between \emph{steps} using the following commands: 

\begin{itemize}
\item \emph{Step Over:} the debugger \blue{steps} over a given line. If the line contains a function, then the function \blue{is} executed, and the result returned without stepping through each of its lines. 

\item \emph{Step Into:} the debugger \blue{enters} the function at the current line and continue stepping from there, line-by-line.

\item \emph{Step Out:} this action would take the debugger back to the line where the current function was called.
\end{itemize}

To start an interactive debugging session, developers set a breakpoint. If not, the IDE would not stop and enter its interactive mode. For example, Eclipse IDE automatically opens the “Debugging Perspective” when execution hits a breakpoint. A developer can run a system in debugging mode without setting breakpoints, but she must set a breakpoint to be able to stop the execution, step in, and observe variable states. Briefly, there is no interactive debugging session without at least one breakpoint \blue{set in the code}.

\noindent Finally, some debuggers allow debugging \textit{remotely}, for example, to perform hot-fixes or to test mobile applications and systems operating in remote configurations.

\subsection{Self-organization and Swarm Intelligence} 

Self-organization is a concept emerged from Social Sciences and Biology and it is defined as the set of dynamic mechanisms enabling structures to appear at the global level of a system from interactions among its lower-level components, without being explicitly coded at the lower levels. Swarm intelligence (SI) describes the behavior resulting from the self-organization of social agents (as insects) \cite{Garnier2007}. Ant nests and the societies that they house are examples of SI \cite{Tschinkel2015}. Individual ants can only perform relatively simple activities, yet the whole colony can collectively accomplish sophisticated activities. Ants achieve SI by exchanging information encoded as chemical signals---pheromones, \eg{} indicating a path to follow or an obstacle to avoid.

Similarly, SI could be used as a metaphor to understand or explain the development of a multiversion large and complex software systems built by software teams. Individual developers can usually perform activities without having a global understanding of the whole system \cite{Ball1996}. In a bird's eye view, software development is analogous to some SI in which groups of agents, interacting locally with one another and with their environment and following simple rules, lead to the emergence of global behaviors previously unknown/impossible to the individual agents. We claim that the similarities between the SI of ant nests and  complex software systems are not a coincidence. Cockburn \cite{Cockburn2006} suggested that the best architectures, requirements, and designs emerge from self-organizing developers, growing in steps and following their changing knowledge, and the changing wishes of the user community, \ie{} a typical example of swarm intelligence.

\begin{figure*}[ht]
	\centering
	\includegraphics[width=0.80\linewidth]{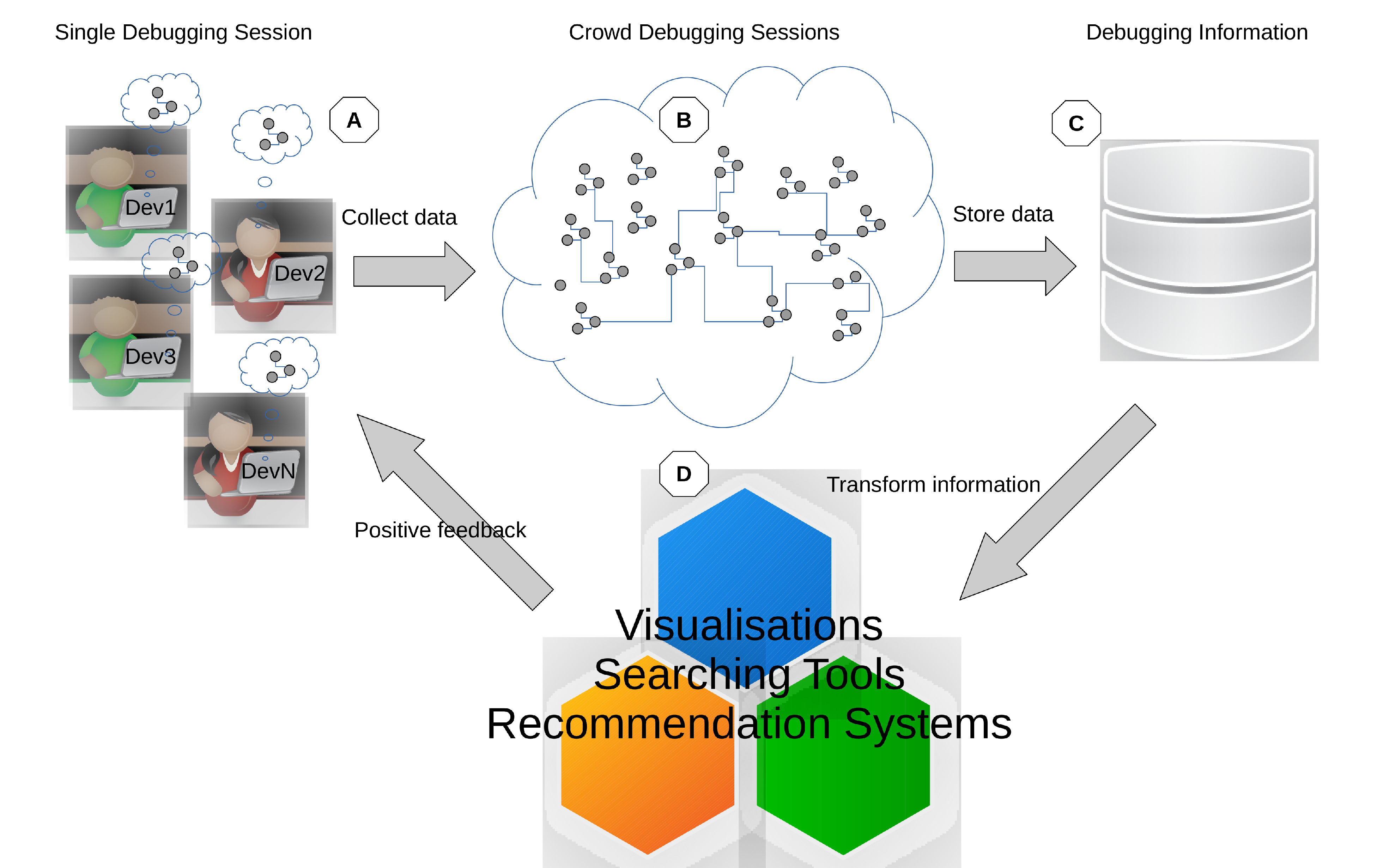}
	\caption{Overview of the \approach approach}
	\label{fig:vision}
\end{figure*}

\subsection{Information Foraging} 

Information Foraging Theory (IFT) is based on the optimal foraging theory developed by Pirolli and Card \cite{Lawrance2013} to understand how people search for information. IFT is rooted in biology studies and theories of how animals hunt for food. It was extended to debugging by Lawrance \ea \cite{Lawrance2013}.

However, no previous work proposes the sharing of knowledge related to debugging activities. Differently from works that use IFT on a model \emph{one prey/one predator} \cite{Piorkowski:2015}, we are interested in \emph{many developers} working independently in many \emph{debugging sessions} and sharing information to allow SI to emerge. Thus, debugging becomes a foraging process in \blue{a SI} environment. 

These concepts---SI and IFT---have led to the design of a crowd approach applied to debugging activities: a different, collective way of doing debugging that collects, shares, retrieves information from (previous and current) debugging sessions to support (current and future) debugging sessions.

%% file: 30SwarmDebugging.tex
\section{The Swarm Debugging Approach}
\label{Section: Our Approach}

Swarm Debugging (SD) uses swarm intelligence applied to interactive debugging data to create knowledge \blue{for supporting} software development activities. \approach works as follows. 

First, several developers perform their individual, independent debugging activities. During \blue{these activities}, debugging events are collected by listeners (Label A in Figure \ref{fig:vision}), for example, breakpoints-toggling and stepping events (Label B in Figure \ref{fig:vision}), that are then stored in a debugging-knowledge repository (Label C in Figure \ref{fig:vision}). \blue{For accessing} this repository, services are defined and implemented in the SDI. For example, stored events are processed by dedicated algorithms (Label D in Figure \ref{fig:vision}) (1) to create (several types of) visualizations, (2) to offer (distinct ways of) searching, and (3) to provide recommendations to assist developers during debugging. Recommendations \blue{are related} to the locations where to toggle breakpoints. Storing and using these events allow \blue{sharing} developers' knowledge among developers, creating a collective intelligence about the software systems and their debugging.

We chose to instrument the Eclipse IDE, a popular IDE, to implement \approach and to reach a large number of users. Also, we use services in the cloud to collect the debugging events, to process these events and to provide visualizations and recommendations from these events. Thus, we decoupled data collection from data usage, allowing other researchers/tools vendors to use the collected data.

During debugging, developers analyze the code, toggling breakpoints and stepping in and through statements. While traditional dynamic analysis approaches collect all interactions, states or events, \acron collects only invocations explicitly explored by developers : SDI collects only visited areas and paths  (chains of invocations by \eg \textit{Step Into} or F5 in Eclipse IDE) and, thus,  does not suffer from performance or memory issues as omniscient debuggers \cite{Pothier2009} or tracing-based approaches could.

Our decision to record information about breakpoints and stepping is well supported by a study from \blue{Beller \al} \cite{beller2018dichotomy}. A finding of this study is that setting breakpoints and stepping through code are the most used debugging features. They showed that most of the recorded debugging events are related to the creation (4,544), removal (4,362) or adjustment of breakpoints, hitting them during debugging and stepping through the source code. Furthermore, other advanced debugging features like defining watches and modifying variable values have been much less used \cite{beller2018dichotomy}.

\section{\blue{SDI in a Nutshell}}

\blue{To evaluate the Swarm Debugging \blue{approach}, we have implemented the \textbf{Swarm Debugging Infrastructure} (see ~\url{https://github.com/SwarmDebugging}). The Swarm Debugging Infrastructure (SDI) \cite{PetrilloQRS2016} provides a set of tools for collecting, storing, sharing, retrieving, and visualizing data collected during developers' debugging activities. The SDI is an Eclipse IDE\footnote{https://www.eclipse.org/} plug-in, integrated with Eclipse Debug core. It is organized in three main modules: (1) the Swarm Debugging Services; (2) the Swarm Debugging Tracer;  and, (3) Swarm Debugging Views. All the implementation details of SDI are available in the \textit{Appendix} section.}

\subsection{\blue{Swarm Debugging Global View}}
\label{sec:sdgv}

\begin{figure}
	\centering
	\includegraphics[width=0.8\linewidth]{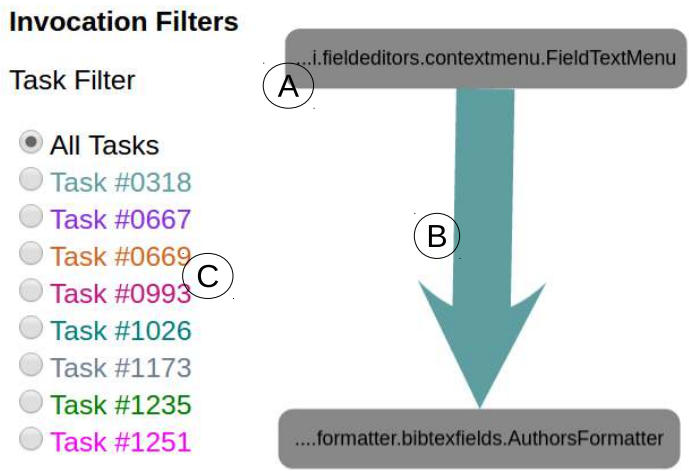}
	\caption{GV elements - Types (nodes), invocations (edge) and Task filter area.}
	\label{fig:SDGVDetails}
\end{figure}

\blue{Swarm Debugging Global View (GV) is a call graph for modeling software based on directed call graph~\cite{Grove1997} to explicit the hierarchical relationship by invocated methods. This visualization use rounded gray boxes (Figure \ref{fig:SDGVDetails}-A) to represent types or classes (nodes) and oriented arrows (Figure \ref{fig:SDGVDetails}-B) to express invocations (edges). GV is built using previous debugging session context data collected by developers for different tasks.}

\begin{figure*}
	\centering
	\includegraphics[angle=0,width=1\linewidth]{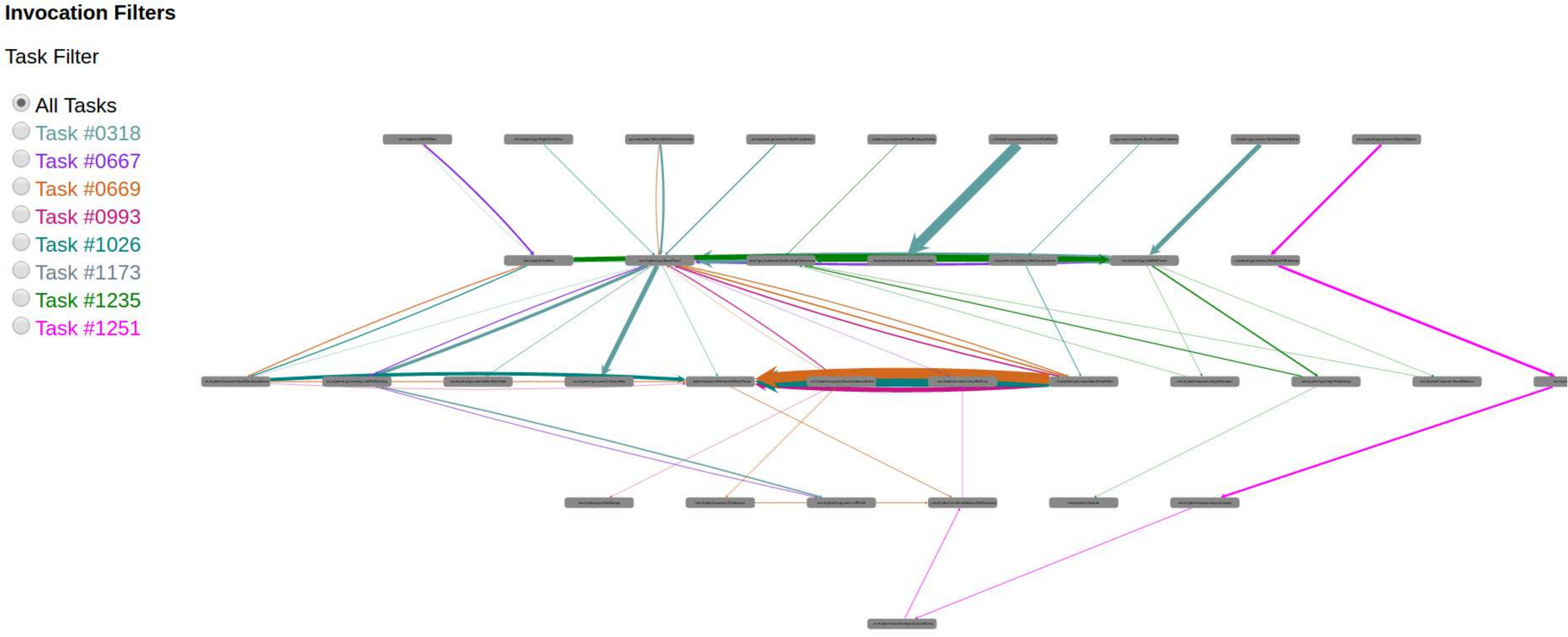}
	\caption{GV on all tasks}
	\label{fig:fullgv}
\end{figure*}

\blue{GV was implemented using CytoscapeJS \cite{Saito2012}, a Graph API JavaScript framework, applying an automatic layout manager \textit{breadthfirst}. As a web application, the SD visualisations can be integrated into an Eclipse view as an SWT Browser Widget, or accessed through a traditional browser such as Mozilla Firefox or Google Chrome.}

\blue{In this view, the grey boxes are types that developers visited during debugging sessions. The edges represent method calls (Step Into or F5 on Eclipse) performed by all developers in all traced tasks on a software project. Each edge colour represents a task, and line thickness is proportional to the number of invocations. Each debugging session contributes with a context, generating the visualisation combining all collected invocations. The visualisation is organised in layers or stacks, and each line is a layer of invocations.  The starting points (non-invoked methods) are allocated on top of a tree, the adjacent nodes in an invocation sequence. Besides, developers can directly go to a type in the Eclipse Editor by double-clicking over a node in the diagram. In the left corner, developers can use radio buttons to filter invocations by task (figure \ref{fig:SDGVDetails}-C), showing the paths used by developers during previous debugging sessions by a task. Finally, developers can use the mouse to pan and zoom in/out on the visualisation. Figure \ref{fig:fullgv} shows an example of GV with all tasks for JabRef system, and we have data about 8 tasks.}

\blue{GV is a contextual visualization that shows \textbf{only the paths explicitly and intentionally visited by developers}, including type declarations and method invocations explored by developers based on their decisions.}

%% file: 40Studies.tex
\section{Using SDI to Understand Debugging Activities}
\label{sec:Using SDI to Understand Debugging Activities}

\blue{The first} benefit of SDI is the fact that it allows for collecting detailed information about debugging sessions. Using this information, researchers can investigate developers behaviors during debugging activities. To illustrate this point, we conducted  two experiments using SDI, to understand developers debugging habits: the times and effort with which they set breakpoints and the locations where they set breakpoints.

Our analysis builds upon three independent sets of observations involving in total three systems. Studies 1 and 2 involved \textit{JabRef}, \textit{PDFSaM}, and \textit{Raptor} as subject systems. We analysed 45 video-recorded debugging sessions, available from our own collected videos (Study 1) and an empirical study performed by Jiang \al~\cite{Jiang2016} (Study 2). 

In this study, we answered the following research questions:

\begin{enumerate}
\item[RQ1:] \RQOne
\item[RQ2:] \RQTwo
\item[RQ3:] \RQThree
\item[RQ4:] \RQFour
\end{enumerate}

In this section, we elaborate more on each of the studies. 

\subsection{Study 1: Observational Study on JabRef}

\subsubsection{Subject System}

To conduct this first study, we selected JabRef\footnote{\url{http://www.jabref.org/}} version 3.2 as subject system.  
This choice was motivated by the fact that JabRef's domain is easy to understand thus reducing any \textit{learning effect}. It is composed of relatively independent packages and classes, \ie{} high cohesion, low coupling, thus reducing the potential commingle effect of \emph{low code quality}.

\subsubsection{Participants}

We recruited eight male professional developers via an Internet-based freelancer service\footnote{\url{https://www.freelancer.com/}}. Two participants are experts, and three are intermediate in Java. Developers self-reported their expertise levels, which thus should be taken with caution. \blue{Also}, we recruited 12 undergraduate and graduate students at Polytechnique Montr\'{e}al to participate in our study. We surveyed all the participants' background information before the study\footnote{Survey available on \url{https://goo.gl/forms/dxCQaBke2l2cqjB42}}. The survey included questions about participants' self-assessment on their level of programming expertise (Java, IDE, and Eclipse), gender, first natural language, schooling level, and knowledge about TDD, interactive debugging and why usually they use a debugger. 
All participants stated that they had experience 
in \textit{Java} and worked regularly with the debugger of Eclipse. 

\subsubsection{Task Description}

We selected five defects reported in the issue-tracking system of JabRef. We \blue{chose} the task of fixing the faults that would potentially require developers to set breakpoints in different Java classes. To ensure this, we manually conducted the debugging ourselves and verified that \blue{for understanding} the root cause of the faults \blue{we had to set at least two breakpoints during our interactive debugging sessions}. Then, we asked participants to find the locations of the faults described in Issues 318, 667, 669, 993, and 1026. Table \ref{Table: Summary Faults JabRef} summarises the faults using their titles from the issue-tracking system.

\begin{table}[ht]
\caption{Summary of the issues considered in JabRef in Study 1}
\label{Table: Summary Faults JabRef}
\begin{tabular}{cl}
\hline
%\textbf{Issue} & \textbf{Summary} \\
Issues & Summaries \\
\hline
%\hline
318 & ``Normalize to Bibtex name format'' \\ 
%\hline
667 & ``hash/pound sign causes URL link to fail'' \\ 
%\hline
669 & ``JabRef 3.1/3.2 writes bib file in a format\\
& that it will not read''\\
%\hline
993 & ``Issues in BibTeX source opens save dialog\\
& and opens dialog \textit{Problem} with parsing entry'\\
& multiple times''\\
%\hline
1026 & ``Jabref removes comments\\ &inside the Bibtex code'' \\
\hline
\end{tabular}
\end{table}

\subsubsection{Artifacts and Working Environment}

We provided the participants with a tutorial\footnote{\url{http://swarmdebugging.org/publication}} explaining how to install and configure the tools required for the study and how to use \blue{them} through a warm-up task. We also presented a video\footnote{\url{https://youtu.be/U1sBMpfL2jc}} to guide the participants during the warm-up task. \blue{In a second document, we} described the five faults and the steps to reproduce them. We also provided participants with a video demonstrating step-by-step how to reproduce the five defects to help them get started. 

We provided a pre-configured Eclipse workspace to the participants and asked them to install Java 8, Eclipse Mars 2 with the Swarm Debugging Tracer plug-in \cite{PetrilloQRS2016} to collect automatically breakpoint-related events. The Eclipse workspace contained two Java projects: a Tetris game for the warm-up task and JabRef v3.2 for the study. We also required that the participants install and configure the Open Broadcaster Software\footnote{\url{https://obsproject.com}} (OBS), open-source software for live streaming and recording. We used the OBS to record the participants' screens.

\subsubsection{Study Procedure}

After installing their environments, we asked participants to perform a warm-up task with a Tetris game. The task consisted of starting a debugging session, setting a breakpoint, and debugging the Tetris program to locate a given method. We used this task to confirm that the participants' environments were properly configured and also to accustom the participants with the study settings. It was a trivial task that we also used to filter the participants who would have too little knowledge of Java, Eclipse, and Eclipse Java debugger. All participants who participated in our study correctly executed the warm-up task.

After performing the warm-up task, each participant performed debugging to locate the faults. We established a maximum limit of one-hour per task and informed the participants that the task \blue{would require} about 20 minutes for each fault, which we will discuss as a possible threat to validity. We based this limit on previous experiences with these tasks during mock trials. After the participants performed each task, we \blue{asked them to answer} a post-experiment questionnaire to collect information about the study, asking if they found the faults, where were the faults, why the faults happened, if they were tired, and a general summary of their debugging experience.

\subsubsection{Data Collection}

\blue{The Swarm Debugging Tracer plug-in automatically and transparently collected all debugging data (breakpoints, stepping, method invocations). Also}, we recorded the participant's screens during their debugging sessions with OBS. We collected the following data:

\begin{itemize}
\item 28 video recordings, one per participant and task, which are essential to control the quality of each session and to produce a reliable and reproducible chain of evidence for our results.

\item The statements (lines in the source code) where the participants set breakpoints. We considered the following types of statements because they are representative of the main concepts in any programming languages:

\begin{itemize}
\item \textit{call}: method/function invocations;
\item \textit{return}: returns of values;
\item \textit{assignment}: settings of values;
\item \textit{if-statement}: conditional statements;
\item \textit{while-loop}: loops, iterations. 
\end{itemize}

\item  Summaries of the results of the study, one per participant, via a questionnaire, which included the following questions:
\begin{itemize}
\item Did you locate the fault?
\item Where was the fault? 
\item Why did the fault happen?
\item Were you tired?
\item How was your debugging experience?
\end{itemize}
\end{itemize}

Based on this data, we obtained or computed the following metrics, per participant and task:

\begin{itemize}
\item Start Time ($ST$): the timestamp when the participant started a task. We analysed each video, and we started to count when effectively the participant started a task, \ie{} when she started the Swarm Debugging Tracer plug-in, for example.
\item Time of First Breakpoint ($FB$): the time when the participant set her first breakpoint.
\item End time ($T$): the time when the participant finished a task.
\item Elapsed End time ($ET$): $ET = T - ST$
\item Elapsed Time First Breakpoint ($EF$): $EF = FB - ST$
\end{itemize}

We \blue{manually verified} whether participants were successful or not at completing their tasks by analysing the answers provided in the questionnaire and the videos. We knew the locations of the faults because all tasks were solved by JabRef's developers, who completed the corresponding reports in the issue-tracking system, with the changes that they made.

\subsection{Study 2: Empirical Study on PDFSaM and Raptor}

The second study consisted of the re-analysis of 20 videos of debugging sessions available \blue{from an empirical study on change-impact analysis with professional developers~\cite{Jiang2016}.} The authors conducted their work in two phases. In the first phase, they asked nine developers to read two fault reports from two open-source systems and to fix these faults. The objective was to observe the developers' behaviour as they fixed the faults. In the second phase, they analysed the developers' behaviour to determine whether the developers used any tools for change-impact analysis and, if not, whether they performed change-impact analysis manually. 

The two systems analysed in their study are PDF Split and Merge\footnote{\url{http://www.pdfsam.org/}} (PDFSaM) and Raptor\footnote{\url{https://code.google.com/p/raptor-chess-interface/}}. They chose one fault report per system for their study. They chose these systems due to their non-trivial size and because the purposes and domains of these systems were clear and easy to understand \cite{Jiang2016}. The choice of the fault reports followed the criteria that they were already solved and that they could be understood by developers who did not know the systems. Alongside each fault report, they presented the developers with information about the systems, their purpose, their main entry points, and instructions for replicating the faults.

\subsection{Results}

\blue{As can be noticed, Studies 1 and 2 have different approaches. The tasks in Study 1 were fault location tasks, developers did not correct the faults, while the ones in Study 2 were fault correction tasks. Moreover, Study 1 explored five different faults while Study 2 only analysed one fault per system. The collected data provide a diversity of cases and allow a rich, in-depth view of how developers set breakpoints during different debugging sessions.}

\blue{In the following, we present the results regarding each research question addressed in the two studies.} 

\subsection*{RQ1: \RQOne}

We normalised  the elapsed time between the start of a debugging session and the setting of the first breakpoint, $EF$, by dividing it by the total duration of the task, $ET$, to compare the performance of participants across tasks (see Equation \ref{equ:normal}).

\begin{equation}
MFB = \frac{EF}{ET}
\label{equ:normal}
\end{equation}

\begin{table}[ht]
    \centering
    \caption{Elapsed time by task (average) - Study 1 (JabRef) and Study 2}
    \begin{tabular}{ l c c }
        \hline 
        Tasks & Average Times (min.) & Std.\ Devs.\ (min.)\\ 
        \hline 
        318 & 44 &64\\
        667 & 28 &29\\
        669 & 22 &25\\
        993 & 25 &25\\
        1026& 25 &17\\ 
        PdfSam&54&18\\ 
        Raptor&59&13\\
        \hline 
    \end{tabular} 
    \label{tab:elapsetime}
\end{table}

Table \ref{tab:elapsetime} shows the average effort (in minutes) for each task. We find in Study 1 that, on average participants spend 27\% of the total task duration to set the first breakpoint (std.\ dev.\ 17\%). In Study 2, it took on average 23\% of the task time to participants to set the first breakpoint (std.\ dev.\ 17\%).  

\resbox{We conclude that the effort for setting the first breakpoint takes near \blue{one-quarter} of the total effort of a single debugging session\footnote{In fact, there is a “debugging task” that starts when a developer starts to investigate the issue to understand and solve it. There is also an “interactive debugging session” that starts when a developer sets their first breakpoint and decides to run an application in “debugging mode”. Also, a developer could need to conclude one debugging task \blue{in} one-to-many interactive debugging sessions.}. So, this effort is important, and this result suggest that debugging time could be reduced by providing tool support for setting breakpoints.}

\subsection*{RQ2: \RQTwo}

\begin{figure*}[ht]
	\centering
	\includegraphics[width=0.9\linewidth]{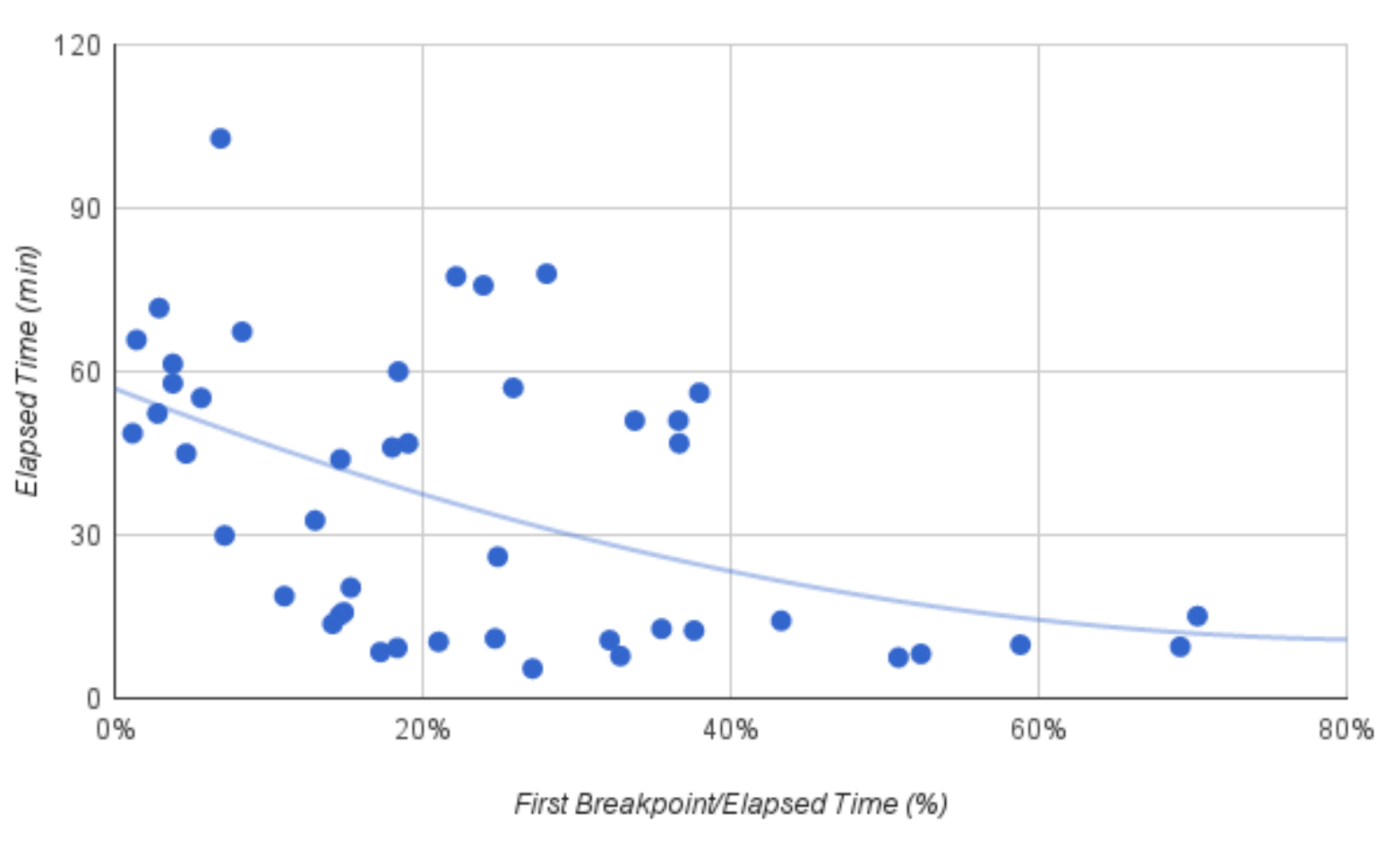}
	\caption{Relation between time of the first breakpoint and task elapsed time (data from the two studies)}
     \label{fig:CombinedStudies}
\end{figure*}

For each session, we normalized the data using Equation \ref{equ:normal} and associated the ratios with their respective task elapsed times. Figure \ref{fig:CombinedStudies} combines the data from the debugging sessions, \blue{each point in the plot represents a debugging session with a specific rate of breakpoints per minute.} Analysing the first breakpoint data, we found a correlation between task elapsed time and time of the first breakpoint ($\rho = -0.47$), resulting that task elapsed time \blue{is inversely correlated} to the time of task's first breakpoint:

\begin{equation}
f(x) = \frac{\alpha}{x^{\beta}}  
\end{equation} 

\noindent where $\alpha = 12$ and $\beta = 0.44$.

\resbox{We observe that when developers toggle breakpoints carefully, they complete tasks faster than developers who set breakpoints quickly.}

This finding also corroborates previous results \blue{found with} a different set of tasks \cite{PetrilloQRS2016}.

\subsection*{RQ3: \RQThree}

We classified the types of statements on which the participants set their breakpoints, and analysed each breakpoint. For Study 1, Table \ref{tab:kinds} shows for example that 53\% (111/207) of the breakpoints are set on \textbf{call statements} while only 1\% (3/207) are set on while-loop statements. For Study 2, Table \ref{tab:kinds2} shows similar trends: 43\% (43/100) of breakpoints are set on \textbf{call statements} and only 4\% (3/207) on while-loop statements. The only difference is on assignment statements, where in Study 1 we found 17\% while Study 2 showed 27\%. After grouping \textit{if-statement}, \textit{return}, and \textit{while-loop} into \textit{control-flow} statements, we found that 30\% of breakpoints are on control-flow statements while \textbf{53\% are on call statements}, and 17\% on assignments.

\begin{table}[ht]
    \centering
    \caption{Study 1 - Breakpoints per type of statement}
    \begin{tabular}{ l  c  c}
        \hline 
        Statements& Numbers of Breakpoints & \%\\ 
        \hline 
        call        &111&53\\ 
        if-statement&39    &19\\ 
        assignment    &36 &17\\ 
        return         &18 &10\\ 
        while-loop    &3  &1\\ 
        \hline 
    \end{tabular} 
    \label{tab:kinds}
\end{table}

\begin{table}[ht]
    \centering
    \caption{Study 2 - Breakpoints per type of statement}
    \begin{tabular}{ l  c  c}
        \hline 
        Statements& Numbers of Breakpoints & \%\\ 
        \hline  
        call        &43&43\\ 
        if-statement&22    &22\\ 
        assignment    &27 &27\\ 
        return         &4 &4\\ 
        while-loop    &4  &4\\ 
        \hline 
    \end{tabular} 
    \label{tab:kinds2}
\end{table}

\resbox{Our results show that in both studies, 50\% of the breakpoints were set on \textit{call} statements while \textit{control-flow} related statements were comparatively fewer, being the \textit{while-loop} statement the least common (2\--4\%)}

\subsection*{RQ4: \RQFour}

We investigated each breakpoint to assess whether there were breakpoints on the same line of code for different participants, performing the same tasks, \ie{} resolving the same fault, by comparing the breakpoints on the same task and different tasks. We sorted all the breakpoints from our data by the \textit{Class} in which they were set and \textit{line number}, and we counted how many times a breakpoint was set on \emph{exactly the same line of code} across participants. We report the results in Table \ref{tab:samelinejabref} for Study 1 and in Tables \ref{tab:samelinepdfsam} and \ref{tab:samelineraptor} for Study 2. 

In Study 1, we found 15 lines of code with two or more breakpoints on the same line for the same task by different participants. In Study 2, we observed breakpoints on exactly the same lines for eight lines of code in PDFSaM and six in Raptor. For example, in Study 1, on line 969 in \textit{Class BasePanel}, participants set a breakpoint on:

\begin{verbatim}
JabRefDesktop.openExternalViewer(metaData(),
   link.toString(), field); 
\end{verbatim}

Three different participants set three breakpoints on that line for issue 667. Tables \ref{tab:samelinejabref}, \ref{tab:samelinepdfsam}, and \ref{tab:samelineraptor} report all recurring breakpoints. These observations show that participants do not choose breakpoints purposelessly, as suggested by Tiarks and Röhm \cite{Tiarks2013}. We suggest that there is an \emph{underlying rationale} on that decision because different participants set breakpoints on exactly the same lines of code.

\begin{table}[ht]
    \centering
    \caption{Study 1 - Breakpoints in the same line of code (JabRef) by task}
    \begin{tabular}{ c l  l  c }
\hline 
Tasks&Classes&Lines of Code&Breakpoints\\
\hline 
%\hline
0318&	AuthorsFormatter	&43		&5\\
0318&	AuthorsFormatter	&131	&3\\
0667&	BasePanel 			&935	&2\\
0667&	BasePanel 			&969	&3\\
0667&	JabRefDesktop 		&430	&2\\
0669&	OpenDatabaseAction 	&268	&2\\
0669&	OpenDatabaseAction 	&433	&4\\
0669&	OpenDatabaseAction 	&451	&4\\
0993&	EntryEditor 		&717	&2\\
0993&	EntryEditor 		&720	&2\\
0993&	EntryEditor 		&723	&2\\
0993&	BibDatabase 		&187	&2\\
0993&	BibDatabase 		&456	&2\\
1026&	EntryEditor 		&1184	&2\\
1026&	BibtexParser 		&160	&2\\
\hline 
\end{tabular} 
\label{tab:samelinejabref}
\end{table}

\begin{table}[ht]
    \centering
    \caption{Study 2 - Breakpoints in the same line of code (PdfSam)}
    \begin{tabular}{ l  l  c }
\hline 
Classes&Lines of Code&Breakpoints\\
\hline 
PdfReader&230&2\\ 
PdfReader&806&2\\ 
PdfReader&1923&2\\ 
ConsoleServicesFacade&89&2\\ 
ConsoleClient&81&2\\
PdfUtility&94&2\\
PdfUtility&96&2\\
PdfUtility&102&2\\ 
\hline 
\end{tabular} 
\label{tab:samelinepdfsam}
\end{table}

\begin{table}[ht]
    \centering
    \caption{Study 2 - Breakpoints in the same line of code (Raptor)}
    \begin{tabular}{ l  l  c }

\hline 
Classes&Lines of Code&Breakpoints\\
\hline 
icsUtils&333&3\\ 
Game&1751&2\\
ExamineController&41&2\\  
ExamineController&84&3\\ 
ExamineController&87&2\\ 
ExamineController&92&2\\ 
\hline
\end{tabular} 
\label{tab:samelineraptor}
\end{table}

When analysing Table \ref{tab:samelinejabrefNoTask}, we found 135 lines of code having two or more breakpoints for different tasks by different participants. For example, five different participants set five breakpoints on the line of code 969 in \textit{Class BasePanel} independently of their tasks (\blue{\textbf{in that case for three different tasks}}). This result suggests a potential opportunity to recommend those locations as candidates for new debugging sessions.

\begin{table}[ht]
    \centering
    \footnotesize
    \caption{Study 1 - Breakpoints in the same line of code (JabRef) in all tasks}
    \begin{tabular}{lll}
\hline
Classes&Lines of Code&Breakpoints\\
\hline
\hline
BibtexParser&138,151,159&2,2,2\\
			&160,165,168&3,2,3\\
			&176,198,199,299&2,2,2,2\\
\hline
EntryEditor&717,720,721&3,4,2\\
		   &723,837,842&2,3,2\\
           &1184,1393&3,2\\
\hline
BibDatabase&175,187,223,456&2,3,2,6\\
\hline
OpenDatabaseAction&433,450,451&4,2,4\\
\hline
JabRefDesktop&40,84,430&2,2,3\\
\hline
SaveDatabaseAction&177,188&4,2\\
\hline
BasePanel&935,969&2,5\\
\hline
AuthorsFormatter&43,131&5,4\\
\hline
EntryTableTransferHandler&346&2\\
\hline
FieldTextMenu&84&2\\
\hline
JabRefFrame&1119&2\\
\hline
JabRefMain&8&5\\
\hline
URLUtil&95&2\\
\hline 
\end{tabular} 
\label{tab:samelinejabrefNoTask}
\end{table}

We also analysed if \blue{the} same class received breakpoints for different tasks. We grouped all breakpoints by \emph{class} and counted how many breakpoints were set on the classes for different tasks, putting ``Yes'' if a type had a breakpoint, producing Table \ref{tab:sametypeDiffentTask}. We also counted the numbers of breakpoints by type, and how many participants set breakpoints on a type. 

For Study 1, we observe that ten classes received breakpoints in different tasks by different participants, resulting in 77\% (160/207) of breakpoints. For example, \texttt{class BibtexParser} had 21\% (44/207) of breakpoints in 3 out of 5 tasks by \emph{13 different participants}. (This analysis only applies to Study 1 because \blue{Study 2 has only one task per system, thus not allowing} to compare breakpoints across tasks.)

\begin{table*}[ht]
\centering
\caption{Study 1 - Breakpoints by class across different tasks}
\begin{tabular}{ l  c  c  c  c  c  c  c }
\hline 
Types&Issue 318&Issue 667&Issue 669&Issue 993&Issue 1026&Breakpoints&Dev.\ Diversities\\ 
\hline 
SaveDatabaseAction&  &    &    Yes&    Yes&    Yes&    7&    2\\ 
BasePanel		  & Yes&    Yes&    Yes&    &    Yes&    14&    7\\ 
JabRefDesktop&         Yes&    Yes&    &    &    &    9&    4\\ 
EntryEditor&         &    &    Yes&    Yes&    Yes&    36&    4\\ 
BibtexParser&       &    &    Yes&    Yes&    Yes&    44&    6\\ 
OpenDatabaseAction& &    &    Yes&    Yes&    Yes&    19&    13\\ 
JabRef&             Yes&    Yes&    Yes&    &    &    3&    3\\
JabRefMain&         Yes&    Yes&    Yes&    Yes&    &    5&    4\\
URLUtil&             Yes&    Yes&    &    &   &    4&    2\\ BibDatabase&         &    &    Yes&    Yes&   Yes&    19&    4\\
\hline 
\end{tabular} 
\label{tab:sametypeDiffentTask}
\end{table*}

\begin{figure*}[ht]
	\centering
    \includegraphics[width=1\linewidth]{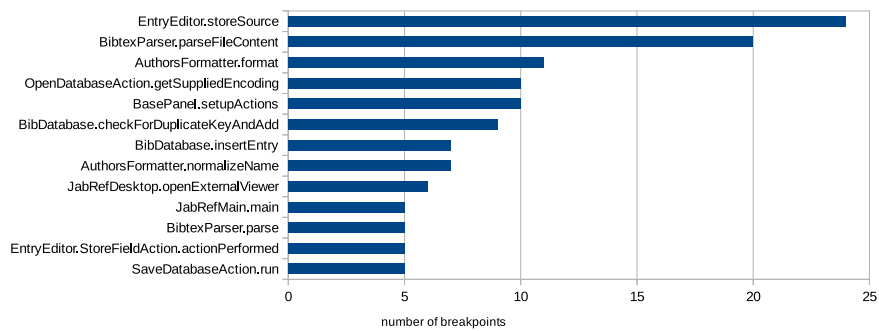}
    \caption{Methods with 5 or more breakpoints}
    \label{fig:methods}
\end{figure*}

Finally, we count how many breakpoints are in the same method across tasks and participants, indicating that there were ``preferred'' methods for setting breakpoints, independently of task or participant. We find that 37 methods received at least two breakpoints, and \emph{13 methods received five or more breakpoints during different tasks by different developers}, as reported in Figure \ref{fig:methods}. In particular, the method \textit{EntityEditor.storeSource} received \emph{24 breakpoints}, and the method \textit{BibtexParser.parseFileContent} received \emph{20 breakpoints} by different developers on different tasks. 

\resbox{\blue{Our results suggest that developers do not choose breakpoints lightly and there is a \textbf{rationale} in their setting breakpoints}, because different developers set breakpoints on the same line of code for the same task, and different developers set breakpoints on the same type or method for different tasks. Furthermore, our results show that different developers, for different tasks, set breakpoints at the same locations. These results show the usefulness of collecting and sharing breakpoints to assist developers during maintenance tasks.}

\section{Evaluation of Swarm Debugging using GV}
\label{sec:Evaluation of SD}

To assess \blue{other} benefits that our approach can bring to developers, we conducted a controlled experiment and interviews focusing on analysing debugging behaviors from 30 professional developers. \blue{We intended} to evaluate if sharing information obtained in previous debugging sessions %involving \textit{JabRef} 
supports debugging tasks.  We wish to answer the following two research questions:

\begin{itemize}
\item[RQ5:] \RQFive
\item[RQ6:] \RQSix
\end{itemize}

\subsection{Study design}

The study consisted of two parts:  (1) a qualitative evaluation using GV in a browser and (2) a controlled experiment on fault location tasks in a Tetris program, using GV integrated into Eclipse. The planning, realization and some results are presented in the following sections.

\subsubsection{Subject System}

For this qualitative evaluation, we chose % again selected 
JabRef\footnote{\url{http://www.jabref.org/}} as subject system. JabRef is a reference management software developed in Java. It is open-source, and its faults are publicly reported. Moreover, JabRef is of reasonably good quality.

\subsubsection{Participants}
\begin{figure}[ht]
	\centering
	\includegraphics[width=0.9\linewidth]{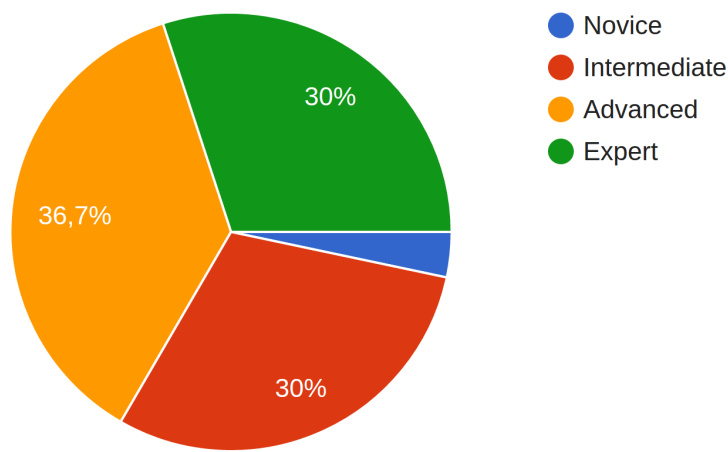}
	\caption{Java expertise}
	\label{fig:java}
\end{figure}

To reproduce a realistic industry scenario, we recruited 30 professional freelancer developers\footnote{\url{https://www.freelancer.com/}}, being 23 male and seven female.  Our participants have on average six years of experience in software development (st.\ dev.\ four years). They have in average 4.8 years of Java experience (st.\ dev.\ 3.3 years), \blue{and 97\% used Eclipse. As shown in Figure \ref{fig:java}, 67\% are advanced or experts on Java.} 

Among these professionals, 23 participated in a qualitative evaluation (qualitative evaluation of GV), and 11 participated in fault location (controlled experiment - 7 control and 6 experiment) using the Swarm Debugging Global View (GV) in Eclipse.

\subsubsection{Task Description}

We chose debugging tasks to trigger the participants' debugging sessions. We asked participants to find the locations of true faults in JabRef. We picked 5 faults reported against JabRef v3.2 in its issue-tracking system, \ie{} Issues 318, 993, 1026, 1173, 1235 and 1251. We asked participants to find the locations of the faults, asking questions as \textit{Where was the fault for Task 318?}, or \textit{For Task 1173, where would you toggle a breakpoint to fix the fault?}, and about positive and negative aspects of GV \footnote{The full qualitative evaluation survey is available on \url{https://goo.gl/forms/c6lOS80TgI3i4tyI2}.}. 

\subsubsection{Artifacts and Working Environment}
After the subject's profile survey, we provided artifacts to support the two phases of our evaluation. For phase one, we provided an electronic form with instructions to follow and questions to answer. The GV was available at  \url{http://server.swarmdebugging.org/}. For phase two, we provided participants with two instruction documents. The first document was an experiment tutorial\footnote{\url{http://swarmdebugging.org/publications/experiment/tutorial.html}} that explained how to install and configure all tools to perform a warm-up task, and the experimental study. We also used the warm-up task to confirm that the participants' environment was correctly configured and that the participants understood the instructions. The warm-up task was described using a video to guide the participants. We make this video available on-line\footnote{\url{https://youtu.be/U1sBMpfL2jc}}. The second document was an electronic form to collect the results and other assessments made using the integrated GV.

For this experimental study, we used Eclipse Mars 2 and Java 8, the SDI with GV and its Swarm Debugging Tracer plug-in, and two Java projects: a small Tetris game for the warm-up task and JabRef v3.2 for the experimental study. All participants received the same workspace, provided by our artifact repository.

\subsubsection{Study Procedure}

The qualitative evaluation consisted of a set of questions about JabRef issues, \blue{using GV on a regular Web browser without accessing the JabRef source code}. We asked the participants to identify the ``type" (classes) in which the faults were located for Issues 318, 667, and 669, using only the GV. We required an explanation for each answer. In addition to providing information about the usefulness of the GV for task comprehension, this evaluation helped the participants to become familiar with the GV.

The controlled experiment was a fault-location task, in which we asked the same participants to find the location of faults using the GV integrated \blue{into} their Eclipse IDE. We divided the participants \blue{into} two groups: a control group (seven participants) and an experimental group (six participants). Participants from the control group performed fault location for Issues 993 and 1026 \textbf{without using the GV} while those from the experimental group \blue{did} the same tasks \textbf{using the GV}.

\subsubsection{Data Collection}

In the qualitative evaluation, the participants answered the questions directly in an electronic form. They used the GV available on-line\footnote{\url{http://server.swarmdebugging.org/}} with collected data for JabRef Issues 318, 667, 669.

In the controlled experiment, each participant executed the warm-up task.
This task consisted in starting a debugging session, toggling a breakpoint, and debugging a Tetris program to locate a given method. After the warm-up task, each participant executed debugging sessions to find the location of the faults described in the five issues. We set a time constraint of one hour. We asked participants to control their fatigue, asking them to go to the next task if they felt tired while informing us of this situation in their reports. Finally, each participant filled a report to provide answers and other information like whether they completed the tasks successfully or not, and (just for the experimental group)  commenting on the usefulness of GV during each task.

All services were available on our server\footnote{\url{http://server.swarmdebugging.org}} during the debugging sessions, and the experimental data were collected \blue{within three} days. We also \blue{captured video from the participants, obtaining more than 3 hours of debugging}. The experiment tutorial contained the instruction to install and set the Open Broadcaster Software \footnote{OBS is available on \url{https://obsproject.com/}.} for video recording tool.

\subsection{Results}

We now discuss the results of our evaluation.

\subsubsection*{RQ5: \RQFive}

During the qualitative evaluation, we asked the participants to analyse the graph generated by GV to identify the type of the location of each fault, \textbf{without reading the task description or looking at the code}. The GV generated graph had invocations collected from previous debugging sessions. These invocations were generated during ``good" sessions (the fault was found\blue{ -- 27/31 sessions}) and ``bad" sessions (the fault was not found \blue{ -- 4/31 sessions}). We analysed results obtained for Tasks 318, 667, and 699, comparing the number of participants who could propose a solution and the correctness of the solutions.

\begin{figure*}[ht]
	\centering
	\includegraphics[width=0.8\linewidth]{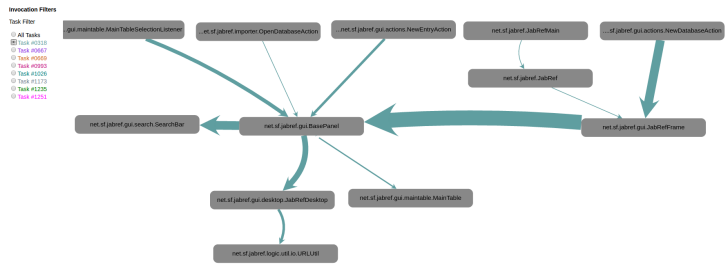}
	\caption{GV for Task 0318}
	\label{fig:GraphTask0318}
\end{figure*}

For Task 318 (Figure \ref{fig:GraphTask0318}), 95\% of participants (22/23) could suggest a ``candidate" type for the location of the fault, just by using the GV view. Among these participants, \textbf{52\% (12/23) suggested correctly \textit{AuthorsFormatter} as the problematic type}.

\begin{figure}[ht]
	\centering
	\includegraphics[width=1\linewidth]{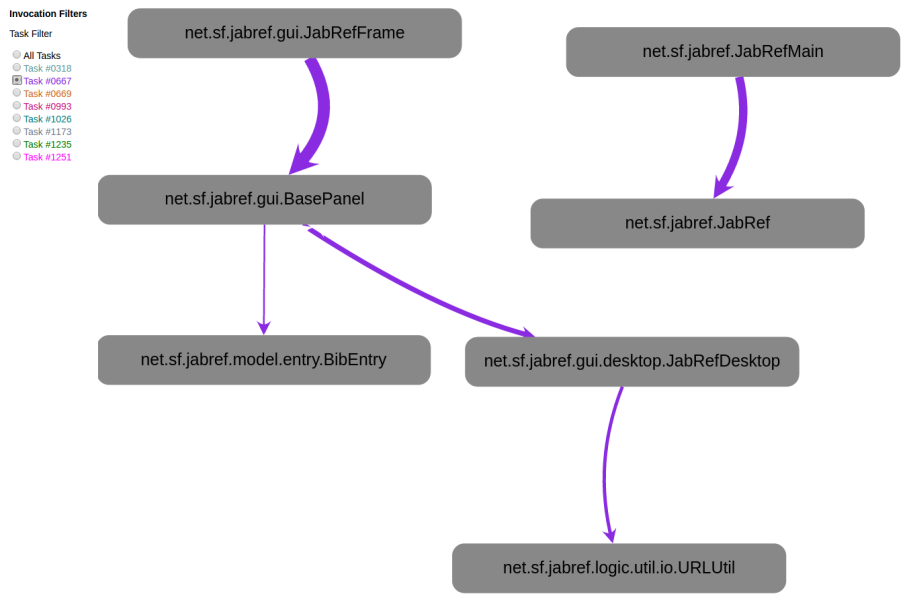}
	\caption{GV for Task 0667}
	\label{fig:GraphTask0667}
\end{figure}

For Task 667 (Figure \ref{fig:GraphTask0667}), 95\% of participants (22/23) could suggest a ``candidate" type for the problematic code, just analysing the graph provided by the GV. Among these participants, \textbf{31\% (7/23) suggested correctly that \textit{URLUtil} was the problematic type}.

\begin{figure}[ht]
	\centering
	\includegraphics[width=1\linewidth]{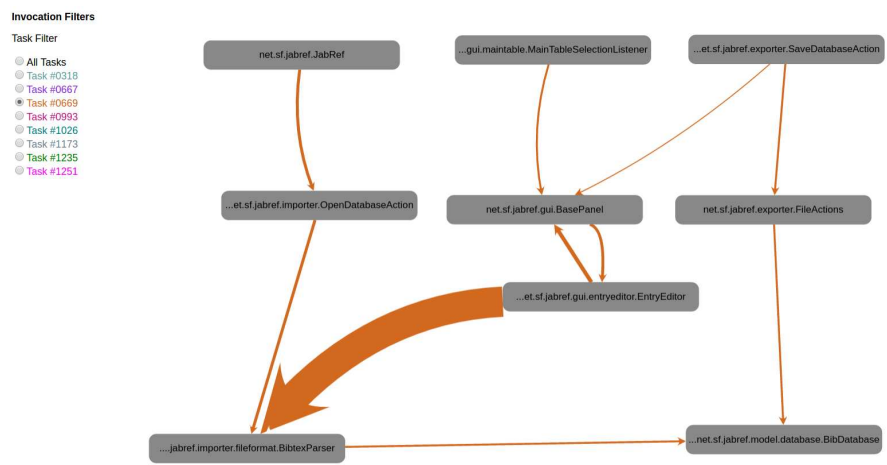}
	\caption{GV for Task 0669}
	\label{fig:GraphTask0669}
\end{figure}

Finally, for Task 669 (Figure \ref{fig:GraphTask0669}), again 95\% of participants (22/23) could suggest a ``candidate" for the type in the problematic code, just by looking at the GV. However, none of them (\ie{} \textbf{0\% (0/23)}) provided the correct answer, which was \textit{OpenDatabaseAction}.

\resbox{Our results show that combining stepping paths in a graph visualisation from several debugging sessions help developers produce correct hypotheses about fault locations without see the code previously.}

\subsubsection*{RQ6: \RQSix}

\begin{figure}[ht]
	\centering
	\includegraphics[width=1\linewidth]{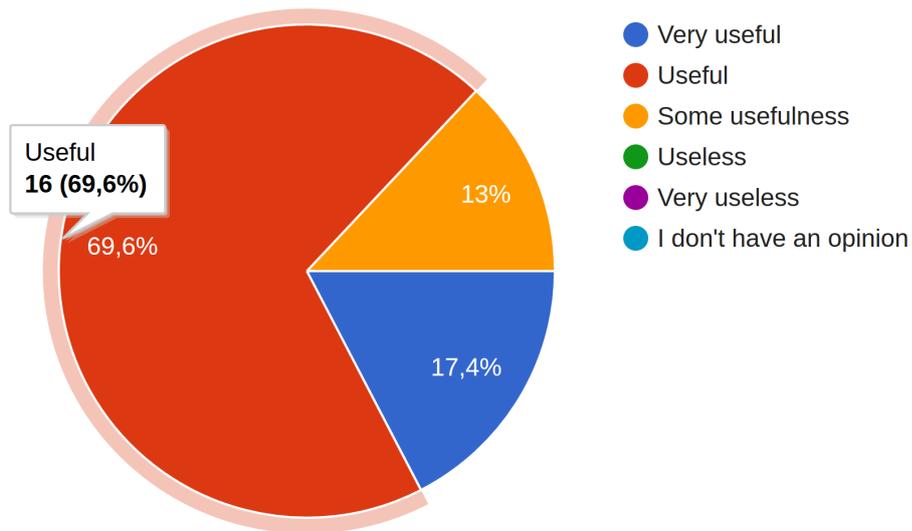}
	\caption{GV usefulness - experimental phase one}
	\label{fig:usefulness1}
\end{figure}

\begin{figure}
	\centering
	\includegraphics[width=1\linewidth]{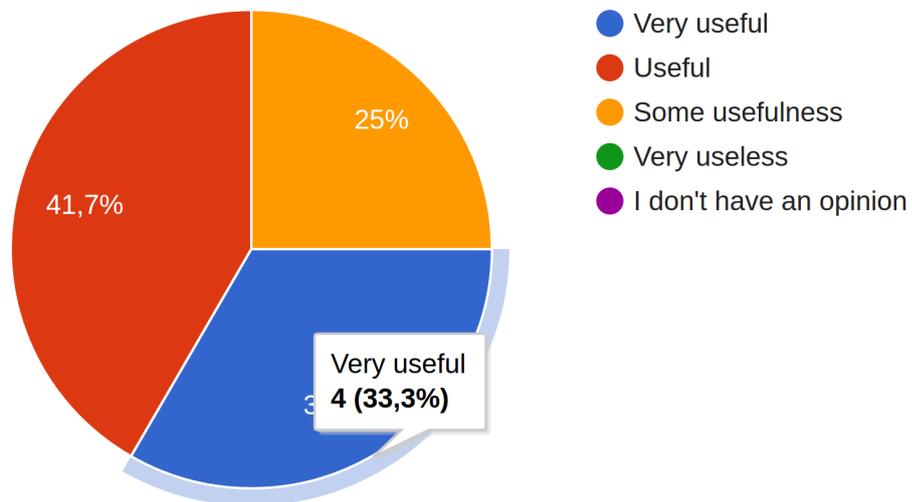}
	\caption{GV usefulness - experimental phase two}
	\label{fig:usefulness2}
\end{figure}

We analysed each video recording and searched for evidence of GV utilisation during fault-locations tasks. Our controlled experiment showed that 100\% of participants of the experimental group used GV to support their tasks (video recording analysis), navigating, reorganizing, and, especially, diving into the type double-clicking on a selected type. We asked participants if GV is useful to support software maintenance tasks. \blue{We report} that \textbf{87\% of participants agreed that GV is useful or very useful (100\% at least \blue{useful})} through our qualitative study (Figure \ref{fig:usefulness1}) and \textbf{75\% of participants claimed that GV is useful or very useful (100\% at least \blue{useful})} on the \blue{task survey after fault-location tasks} (Figure \ref{fig:usefulness2}). Furthermore, several participants' feedback supports our answers.

\blue{The analysis of our results} suggests that GV is useful to support software-maintenance tasks.

\resbox{Sharing previous debugging sessions supports debugging hypotheses and, consequently, reduces the effort on searching of code.}

\subsection{\blue{Comparing Results from the Control and Experimental Groups}}

\blue{We compared the control and experimental groups using three metrics: (1) the time for setting the first breakpoint; (2) the time to start a debugging session; and, (3) the elapsed time to finish the task. We analysed recording sessions of Tasks 0993 and 1026, compiling the average results from the two groups in Table \ref{tab:controlvsexperimentTask}.}

\blue{Observing the results in Table \ref{tab:controlvsexperimentTask}, we observed that the experimental group spent more time to set the first breakpoint (26\% more time for Task 0993 and 77\% more time for Task 1026). The times to start a debugging session are near the same (12\% more time for Task 0993 and 18\% less time for Task 1026) when compared to the control group. However, participants who used our approach \textbf{spent less time to finish both tasks} (47\% less time to Task 0993 and 17\% less time for Task 1026). This result suggests that participants invested more time to toggle carefully the first breakpoint but consecutively completed the tasks faster than participants who toggled breakpoints quickly, corroborating our results in RQ2.}

\resbox{\blue{Our results show that participants who used the  shared debugging data invested more time to decide the first breakpoint but \textbf{reduced their time to finish the tasks}. These results suggest that sharing debugging information using Swarm Debugging can reduce the time spent on debugging tasks.}}

\begin{table*}
\centering
\caption{\blue{Results from control and experimental groups (average)}}
\label{tab:controlvsexperimentTask}
\begin{tabular}{lrrrr}
Task 0993&&&\\
\hline
Metric                  & Control [C]   & Experiment [E]    & $\Delta$ [C-E] (s)  & \% [E/C]\\
\hline
First breakpoint        & 00:02:55      & 00:03:40	        & -44               & 126\%\\
Time to start           & 00:04:44	    & 00:05:18	        & -33               & 112\%\\
\textbf{Elapsed time}	& 00:30:08	    & 00:16:05	        & 843               &  \textbf{53}\%\\
\hline
\\
Task 1026&&&\\
\hline
Metric                  & Control [C]   & Experiment [E]    & $\Delta$ [C-E] (s) & \% [E/C]\\
\hline
First breakpoint        & 00:02:42	    & 00:04:48	        & -126               & 177\%\\
Time to start           & 00:04:02	    & 00:03:43	        &   19               & 92\%\\
\textbf{Elapsed time}	& 00:24:58	    & 00:20:41	        &  257               & \textbf{83}\%\\
\hline
\end{tabular}
\end{table*}

\subsection{Participants' Feedback}

As with any visualisation technique proposed in the literature, ours is a proof of concept with both intrinsic and accidental advantages and limitations. Intrinsic advantages and limitations pertain to the visualisation itself and our design choices, while accidental advantages and limitations concern our implementation. 
During our experiment, we collected the participants’ feedback about our visualisation and now discuss both its intrinsic and accidental advantages and limitations as reported by them. We go back to some of the limitations in the next section that describes threats to the validity of our experiment. We also report feedback from three of the participants.

\subsubsection{Intrinsic Advantage}

\paragraph{Visualisation of Debugging Paths} Participants com\-men\-ded our visualisation for presenting useful information related to the classes and methods followed by other developers during debugging. In particular, one participant reported that ``[i]t seems a fairly simple way to visualize classes and to demonstrate how they interact.'', which comforts us in our choice of both the visualisation technique (graphs) and the data to display (developers’ debugging paths).

\paragraph{Effort in Debugging} Three participants also mentioned that our visualisation shows where developers spent their debugging effort and where there are understanding “bottlenecks”. In particular, one participant wrote that our visualisation “allows the developer to skip several steps in debugging, knowing from the graph where the problem probably comes from."

\subsubsection{Intrinsic Limitations}

\paragraph{Location} One participant commented that “the location where [an] issue occurs is not the same as the one that is responsible for the issue.” We are well aware of this difference between the location where a fault occurs, for example, a null-pointer exception, and the location of the source of the fault, for example, a constructor where the field is not initialised." 

However, we build our visualisation on the premise that developers can share their debugging activities for that particular reason: by sharing, they could readily identify the source of a fault rather than only the location where it occurs. We plan to perform further studies to assess the usefulness of
our visualisation to validate (or not) our premise.

\paragraph{Scalability} Several participants commented on the possible lack of scalability of our visualisation. Graphs are well known to be not scalable, so we are expecting issues with larger graphs \cite{Pienta2015}. Strategies to mitigate these issues include graph sampling and clustering. We plan to add these features in the next release of our technique.

\paragraph{Presentation} Several participants also commented on the (relative) lack of information brought by the visualisation, which is complementary to the limitation in scalability. 

One participant commented on the difference between the graph showing the developers’ paths and the relative importance of classes during execution. Future work should seek to combine both information on the same graph, possibly by combining size and colours: size could relate to the developers’ paths while colours could indicate the ``importance" of a class during execution.

\paragraph{Evolution} One participant commented that the graph is relevant for one version of the system but that, as soon as some changes are performed by a developer, the paths (or parts thereof) may become irrelevant. 

We agree with the participant and accept this limitation because our visualisation is currently implemented for one version. We will explore in future work how to handle evolution by changing the graph as new versions are created.

\paragraph{Trap} One participant warned that our visualisation could lead developers into a “trap” if
all developers whose paths are displayed followed the “wrong” paths. We agree with the participant but accept this limitation because \blue{developers can always choose appropriate paths}.

\paragraph{Understanding} One participant reported that the visualisation alone does not bring enough information to understand the task at hand. We accept this limitation because
our visualisation is built to be complementary to other views
available in the IDE.

\subsubsection{Accidental Advantages}

\paragraph{Reducing Code Complexity} One participant discussed the use of our visualisation to reduce code complexity for the developers by highlighting its main functionalities.

\paragraph{Complementing Differential Views} Another participant contrasted our visualisation with Git Diff and mentioned
that they complement each other well because our visualisation “[a]llows to quickly see where the problem probably has been before it got fixed.” while Git Diff allows \blue{seeing} where the problem was fixed.

\paragraph{Highlighting Refactoring Opportunities} A third participant suggested that the larger node could represent  classes that could be refactored if they also have many faults, to simplify future debugging sessions for developers.

\subsubsection{Accidental Limitations}

\paragraph{Presentation} Several participants commented on the presentation of the information by our visualisation. Most
importantly, they remarked that identifying the location of the fault was difficult because there was no distinction between faulty and non-faulty classes. In the future, we will assess the use of icons and–or colours to identify faulty classes/methods. 

Others commented on the lack of captions describing the various visual elements. Although this information was present in the tutorial and questionnaires, we will add it also into the
visualisation, possibly using tooltips. 

One participant added that more information, such as “execution time metrics [by] invocations” and “failure/success rate [by] invocations” could be valuable. We plan to perform other controlled experiments with such additional information to \blue{assess} its impact on developers’ performance. 

Finally, one participant mentioned that arrows would sometimes overlap, which points to the need \blue{for} a better layout algorithm for the graph in our visualisation. However, \blue{finding a good} graph layout is a well-known difficult problem.

\paragraph{Navigation} One participant commented that the visualisation does not help developers navigating between classes whose methods have low cohesion. It should be possible to show in different parts of the graph the methods and their classes independently to avoid large nodes. We plan to modify the graph visualisation to have a “method-level” view whose nodes could be methods and–or clusters of methods (independently of their classes).

\subsubsection{General Feedback}

Three participants left general feedback regarding their experience with our visualisation under the question “Describe your debugging experience”. All three participants provided positive comments. We report herein one of the three comments:

\begin{quote}
\textit{It went pretty well. In the beginning I was at a loss, so just was looking around for some time. Then I opened the breakpoints view for another task that was related to file parsing in the hope to find some hints. And indeed I’ve found the BibtexParser class where the method with the most number of breakpoints was the one where I later found the fault. However, only this knowledge was not enough, so I had to study the code a bit. Luckily, it didn’t require too much effort to spot the problem because all the related code was concentrated inside the parser class. Luckily I had a BibTeX database at hand to use it for debugging. It was excellent.}
\end{quote}

This comment highlights the advantages of our approach and %confirms
suggests that our premise may be correct and that developers may benefit from one another’s debugging sessions. It encourages us to pursue our research work in this direction and perform more experiments \blue{to point further ways of improving} our approach.

%% file: 60Discussion.tex
\section{Discussion}
\label{Section: Discussions}

We now discuss some implications of our work for Software Engineering researchers, developers, debuggers' developers, and educators. SDI (and GV) is \blue{open and freely available on-line\footnote{\url{http://github.com/swarmdebugging}}, and}  researchers can use them to perform new empirical studies about debugging activities.

\textbf{Developers can use SDI to record their debugging patterns} to identify debugging strategies that are more efficient in the context of their projects to improve their debugging skills.

\textbf{Developers can share their debugging activities}, such as breakpoints and--or \blue{stepping paths,} to improve collaborative work and ease debugging. While developers usually work on specific tasks, there are sometimes re-open issues and--or similar tasks that need to understand or toggle breakpoints on the same entity. Thus, using breakpoints previously toggled by a developer could help to assist another developer working on a similar task. For instance, the breakpoint search tools can be used to retrieve breakpoints from previous debugging sessions, which could help speed up a new \blue{one}, providing developers with valid starting points. Therefore, the breakpoint searching tool can decrease the time spent to toggle a new breakpoint.

\textbf{Developers of debuggers can use SDI to understand developers' debugging habits} to create new tools -- using novel data-mining techniques -- to integrate different data sources. SDI provides a transparent framework for developers to share debugging information, creating a collective intelligence about their projects.

\textbf{Educators can leverage SDI to teach interactive debugging techniques}, tracing their students' debugging sessions, and evaluating their performance. Data collected by SDI \blue{from debugging sessions performed by} professional developers could also be used to educate students, \eg{} by showing them examples of good and bad debugging patterns.

\blue{There are locations (line of code, class, or method) on which there were set many breakpoints in different tasks by different developers, and this is} an opportunity to recommend those locations as candidates for new debugging sessions. However, we could face the bootstrapping problem: we cannot know that these locations are important until developers start to put breakpoints on them. This problem could be addressed with time, by using the infrastructure to collect and share breakpoints, accumulating data that can be used for future debugging sessions. Further, \blue{such incremental} usefulness can encourage more developers to collect and share breakpoints, possibly leading to better-automated recommendations. 

We \blue{have answered} what debugging information is useful to share among developers to ease debugging with evidence that sharing debugging breakpoints and sessions \blue{can} ease developers' debugging activities. Our study provides useful insights to researchers and tool developers on how to provide appropriate support during debugging activities in general: they could support developers by sharing other developers' breakpoints and sessions. They could also develop recommender systems to help developers in deciding where to set breakpoints\blue{,and} use this evidence to build a grounded theory on the setting of breakpoints and stepping by developers to improve debuggers and other tool support.

%% file: 70Threats.tex
\section{Threats to Validity}
\label{sec:threatsValidity}

Despite its promising results, there exist threats to the validity of our study that we discuss in this section.

As \blue{any other empirical study, ours} is subject to limitations that threaten the validity of its results. The first limitation \blue{is related to the number of participants we had}. With 7 participants, we can not claim generalization of the results. However, we accept this limitation because the goal of the study was to show the effectiveness of the data collected by the SDI to obtain insights about developers' debugging activities. Future studies with \blue{a more significant} number of participants and more systems and tasks are needed to confirm the results of the \blue{present research}.

Other threats to the validity of our study concern their internal, external, and conclusion validity. We accept these threats because the experimental study aimed to show the effectiveness of the SDI to collect and share data about developers' interactive debugging activities. Future work is needed to perform in-depth experimental studies with these research questions and others, possibly drawn from the ones that developers asked \blue{in another study by} Sillito \al{} \cite{Sillito08-QuestionProgrammersAsk}.

\textbf{Construct Validity Threats} are related to the metrics used to answer our research questions. We mainly used breakpoint locations, which is a precise measure. Moreover, as we located breakpoints using our Swarm Debugging Infrastructure (SDI) and visualisation, any issue with \blue{this measure} would affect our results. To mitigate these threats, we collected both SDI data and video captures of the participants' screens and compared the information extracted from the videos with the data collected by the SDI. We observed that the breakpoints collected by the SDI are exactly those toggled by the participants.

We ask participants to self-report on their efforts during the tasks, levels of experience, etc.\ through questionnaires. Consequently, it is possible that the answer does not represent their \emph{real} efforts, levels, etc. We accept this threat because questionnaires are the best means to collect data about participants without incurring a high cost. Construct validity could be improved in future work by using instruments to measure effort independently, for example, but \blue{this} would lead to more time- and effort-consuming experiments.

\textbf{Conclusion Validity Threats} concern the relations found between independent and dependent variables. In particular, they concern the assumptions of the statistical tests performed on the data and how diverse is the data. We did not perform any statistical analysis to answer our research questions, so our results do not depend on any statistical assumption.

\textbf{Internal Validity Threats} \blue{are related} to the tools used to collect the data and the subject systems, and if the collected data is sufficient to answer the research questions. We collected data using our visualisation. We are well aware that our visualisation does not scale for large systems but, for JabRef, it allowed participants to share paths during debugging and researchers to collect relevant data, including shared paths. We plan to \blue{revise} our visualisation \blue{in the near future} to identify possibilities to improve it so that it scales up to large systems.

Each participant performed more than one task on the same system. It is possible that a participant may have \blue{become} familiar with the system after \blue{executing a} task and would be knowledgeable enough to toggle breakpoints when performing the subsequent \blue{ones}. However, we did not observe any significant difference in performance when comparing \blue{the results for the same participant} for the first and last task. Therefore, we accept this threat but still plan for future studies with more tasks on more systems. The participants probably were aware of the fact that all faults were already solved in Github. We controlled this issue using the video recordings, observing that all participants did not look at the commit history during the experiment. 

\textbf{External Validity Threats} are about the possibility to generalise our results. We use only one system in our Study 1 (JabRef) because we needed to have enough data points from a single system to assess the effectiveness of breakpoint prediction. We should collect more data on other systems and check whether the system used can affect our results.

%% file: 80RelatedWork.tex
\section{Related work}
\label{sec:relatedwork}

We now summarise works related to debugging to allow better positioning of our study among the published research. 

\paragraph{Program Understanding} \label{sec:comprehension} Previous work studied program comprehension and provided tools to support program comprehension. Maalej \al{} \cite{Maalej14-Program-Comprehension} observed and surveyed developers during program comprehension \blue{activities}. They concluded that developers need runtime information and reported that developers frequently execute programs using a debugger. Ko \al{} \cite{Ko06-Developers-Find-RelevantInformation} observed that developers spend large amounts of times navigating between program elements. 

Feature and fault location approaches are used to identify and recommend program elements that are relevant to a task at hand \cite{Wang2014}. These approaches use defect report \cite{Zhou2012}, domain knowledge \cite{Ye2014}, version history and defect report similarity \cite{Wang2014} while others, like Mylyn \cite{Kersten06-TaskContext-Productivity}, use developers' interaction traces, which have been used to study work interruption \cite{Sanchez15-Interruption-InteractionHistory}, editing patterns \cite{Ying11-InfuenceOfTask-ProgrammerBehaviour,Zhang12-Mylyn-FileEditingStyles}, program exploration patterns \cite{Soh13-Exploration-Strategy}, or copy/paste behaviour \cite{Ahmed15-Copy-Paste-Behaviour}.

Despite sharing similarities (tracing developer events in an IDE), \blue{our approach differs from Mylyn's~\cite{Kersten06-TaskContext-Productivity}.} First, \blue{Mylyn's approach} does not collect or use any dynamic debugging information; it is not designed to explore the dynamic behaviours of developers during debugging sessions. Second, \blue{it} is useful in editing mode, because it just filters files in an Eclipse view following a previous context. Our approach is for editing mode (finding breakpoints or visualize paths) as during interactive debugging sessions. Consequently, our \blue{work} and Mylyn's are complementary, and they should be used together during development sessions.

\paragraph{Debugging Tools for Program Understanding} 
\label{sec:tools} 

Romero \al{}~\cite{Romero2007} extended the work by Katz and Anderson~\cite{Katz1987} and identified high-level debugging strategies, \eg{} stepping and breaking execution paths and inspecting variable values. They reported that developers use the information available in the debuggers differently depending on their background and level of expertise. 

\textit{DebugAdvisor} \cite{Ashok2009} is a recommender system to improve debugging productivity by automating the search for similar issues from the past. 

Zayour~\cite{Zayour2016} studied the difficulties faced by developers when debugging in IDEs and reported that the features of the IDE affect the times spent by developers on debugging activities.

\paragraph{Automated debugging tools} Automated debugging tools require both successful and failed runs and do not support programs with interactive inputs \cite{Ko2006}. Consequently, they have not been widely adopted in practice. Moreover, automated debugging approaches are often unable to indicate the ``true" locations of faults \cite{Rossler2012}. Other more interactive methods, such as slicing and query languages, help developers but, to date, there has been no evidence that they significantly ease developers' debugging activities. 

Recent studies showed that empirical evidence of the usefulness of many automated debugging techniques is limited~\cite{Parnin2011}. Researchers also found that automated debugging tools are rarely used in practice~\cite{Parnin2011}. At least in some scenarios, the time to collect coverage information, manually label the test cases as failing or passing, and run the calculations may exceed the actual time saved \blue{by} using the automated debugging tools.

\paragraph{Advanced Debugging Approaches} Zheng \al{}~\cite{Zheng2006} presented a systematic approach to the \textit{statistical debugging} of programs in the presence of multiple faults, using probability inference and common voting framework to accommodate more general faults and predicate settings. 
Ko and Myers \cite{Ko2006,Ko2009} introduced \textit{interrogative debugging}, a process with which developers ask questions about their programs outputs to determine what parts of the programs to understand. 

Pothier and Tanter \cite{Pothier2009} proposed \textit{Omniscient debuggers}, \blue{an} approach to support \blue{back-in-time} navigation across previous program states. \emph{Delta debugging} \cite{Hofer2012} by Hofer \al{} means that the smaller the failure-inducing input, the less program code is covered. It can be used to minimise a failure-inducing input systematically. Ressia \cite{Ressia2012} proposed \emph{object-centric debugging}, focusing on objects as the key abstraction execution for many tasks. 

Estler \al{} \cite{Estler2013} discussed \textit{collaborative debugging} suggesting that \blue{collaboration in debugging activities} is perceived as important by developers and can improve their experience. \blue{Our approach is consistent with this finding} although we use asynchronous debugging sessions.

\paragraph{Empirical Studies on Debugging}
Jiang \al~\cite{Jiang2016} studied the change impact analysis process that should be done during software maintenance by developers to make sure changes do not introduce new faults. They conducted two studies about change impact analysis during debugging sessions. They found that the programmers in \blue{their} studies did static change impact analysis before they made changes by using IDE navigational functionalities. \blue{They also} did dynamic change impact analysis after they made changes by running the programs. \blue{In their study,} programmers did not use any change impact analysis tools.

Zhang \al~\cite{Zhang2013} proposed a method to generate breakpoints based on existing fault localization techniques, showing that the generated breakpoints can usually save some human effort for debugging.

%% file: 90Conclusion.tex
\section{Conclusion}
\label{sec:conclusion}

Debugging is an important and challenging task in software maintenance, requiring dedication and expertise. However, despite its importance, developers' debugging behaviors have not been extensively and comprehensively studied. In this paper, we introduced the concept of Swarm Debugging \blue{based on the fact that} developers, performing different debugging sessions build collective knowledge. We asked what debugging information is useful to share among developers to ease debugging. We particularly studied two pieces of debugging information: breakpoints (and their locations) and sessions (debugging paths), because these pieces of information \blue{are related} to the two main activities during debugging: setting breakpoints and stepping in/over/out statements.

To evaluate the usefulness of Swarm Debugging and the sharing of debugging data, we conducted two observational studies. In the first study, to understand how developers set breakpoints, we collected and analyzed more than 10 hours of developers' videos in 45 debugging sessions performed by 28 different, independent developers, containing 307 breakpoints on three software systems. 

\blue{The first study allowed us to draw four main conclusions. At first, setting the first breakpoint is not an easy task and developers need tools to locate the places where to toggle breakpoints. Secondly, the time of setting the first breakpoint is a predictor for the duration of a debugging task independently of the task.  Third, developers choose breakpoints purposefully, with an \emph{underlying rationale}, because different developers set breakpoints on the same line of code for the same task, and  also, different developers toggle breakpoints on the same classes or methods for different tasks, showing the existence of important ``debugging hot-spots" (i.e., regions in the code where there is more incidence of debugging events) and--or more error-prone classes and methods. Finally and surprisingly, different, independent developers set breakpoints at the same locations for similar debugging tasks and, thus, collecting and sharing breakpoints could assist developers during debugging task.} 

Further, we \blue{conducted} a qualitative study with 23 professional developers and a controlled experiment with 13 professional developers, collecting more than 3 hours of developers' debugging sessions. \blue{From this second study, we concluded} that: (1) combining stepping paths in a graph visualisation from several debugging sessions produced elements to support developers' hypotheses about fault locations without looking at the code previously; and (2) \blue{sharing}  previous debugging sessions support debugging hypothesis, and consequently reducing the effort on searching of code. 

\blue{Our results provide evidence that previous debugging sessions provide insights to and can be starting points for developers when building debugging hypotheses. They showed that developers construct correct hypotheses on fault location when looking at graphs built from previous debugging sessions. Moreover, they showed that developers can use past debugging sessions to identify starting points for new debugging sessions. Furthermore, faults are recurrent and may be reopened sometime months later. Sharing debugging sessions (as Mylyn for editing sessions) is an approach to support debugging hypotheses and to support the reconstruction of the complex mental model processes involved in debugging. However, research work is in progress to corroborate these results.}

In future work, we plan to  build grounded theories on the use of breakpoints by developers. We will use these theories to recommend breakpoints to other developers. Developers need tools to locate adequate places to set breakpoints in their source code. Our \blue{results} suggest the opportunity for a breakpoint recommendation system, similar to previous work \cite{Zhang2013}. They could also form the basis for building a grounded theory of the developers' use of breakpoints to improve debuggers and other tool support. 

Moreover, we also suggest that \textbf{debugging tasks could be divided into two activities}, one of \textbf{locating bugs}, which could benefit from the collective intelligence of other developers and could be performed by dedicated ``hunters'', and another one of \textbf{fixing the faults}, which requires deep understanding of the program, its design, its architecture, and the consequences of changes. This latter activity could be performed by dedicated ``builders''. Hence, actionable results include recommender systems and a change of paradigm in the debugging of software programs.

Last but not least, the research community can leverage the SDI to conduct more studies to improve our understanding of developers' debugging \blue{behaviour}, which could ultimately result into the development of whole new families of debugging tools that are more efficient and--or more adapted to the particularity of debugging. Many open questions remain, and this paper is just a first step towards fully understanding how collective intelligence could improve debugging activities.

Our vision is that IDEs should incorporate a general framework to capture and exploit IDE interactions, creating an ecosystem of context-aware applications and plugins. Swarm Debugging is the first step towards intelligent debuggers and IDEs, context-aware programs that monitor and reason about how developers interact with them, providing for crowd software-engineering.

%% file: 95Appendix.tex
\section*{Appendix - Implementation of Swarm Debugging}

\subsection*{Swarm Debugging Services}

\begin{figure}[ht]
	\centering
	\includegraphics[width=0.7\linewidth]{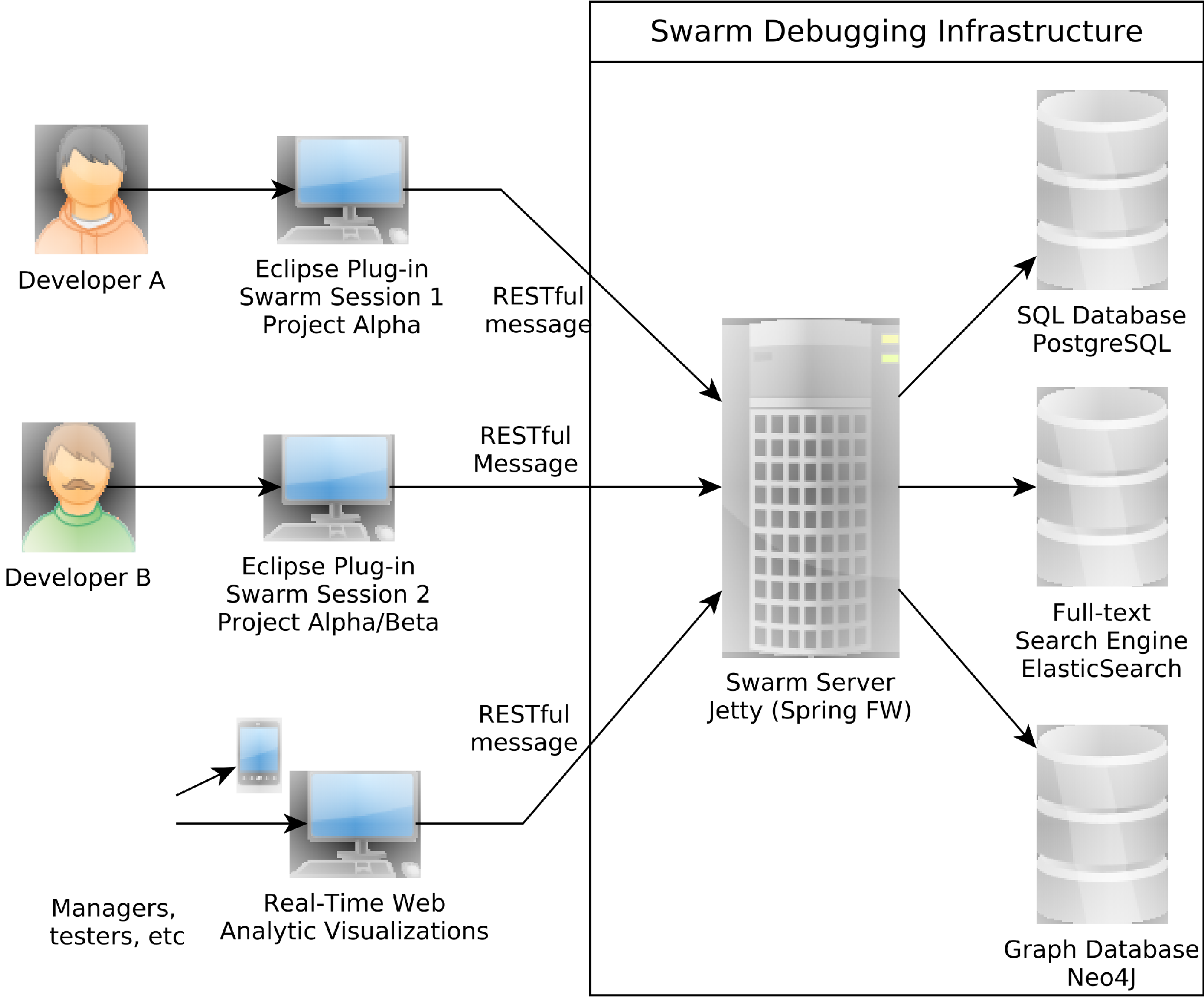}
	\caption{The Swarm Debugging Services architecture}
	\label{fig:infrastructure}
\end{figure}

The Swarm Debugging Services (SDS) provide the infrastructure needed by the Swarm Debugging Tracer (SDT) to store and, later, share debugging data from and between developers. 
Figure \ref{fig:infrastructure} shows the architecture of this infrastructure. 
The SDT sends RESTful messages that are received by a SDS instance that stores them in three specialized persistence mechanisms: an SQL database (PostgreSQL), a full-text search engine (ElasticSearch), and a graph database (Neo4J).

The three persistence mechanisms use similar sets of concepts to define the semantics of the SDT messages. 

We choose and define domain concepts to model software projects and debugging data. Figure \ref{fig:relational} shows the meta-model of these concepts using an entity-relationship representation. The concepts are inspired by the FAMIX Data model \cite{Demeyer1999}. The concepts include:

\begin{figure}[ht]
	\centering
	\includegraphics[width=0.75\linewidth]{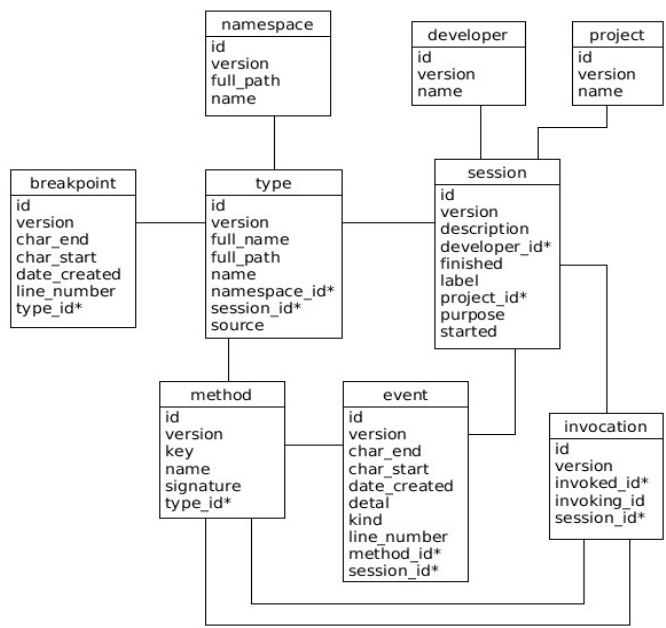}
	\caption{The Swarm Debugging metadata \cite{PetrilloQRS2016}}
	\label{fig:relational}
\end{figure}

\begin{itemize}
\item \textbf{Developer} is a SDT user. She creates and executes debugging sessions.
\item \textbf{Product} is the target software product. A product is a set of Eclipse projects (1 or more).
\item \textbf{Task} is the task to be executed by developers. 
\item \textbf{Session} represents a Swarm Debugging session. It relates developer, project, and debugging events.
\item \textbf{Type} represents classes and interfaces in the project. Each type has a source code and a file. SDS only considers types that have source code available as belonging to the project domain.
\item \textbf{Method} is a method associated with a type, which can be invoked during debugging sessions.
\item \textbf{Namespace} is a container for types. In Java, namespaces are declared with the keyword \textit{package}.
\item \textbf{Invocation} is a method invoked from another method (or from the JVM, in case of the \texttt{main} method).
\item \textbf{Breakpoint} represents the data collected when a developer toggles a breakpoint in the Eclipse IDE. Each breakpoint is associated with a type and a method if appropriate.
\item \textbf{Event} is an event data that is collected when a developer performs some actions during a debugging session.
\end{itemize}

The SDS provides several services for manipulating, querying, and searching collected data: (1) Swarm RESTful API; (2) SQL query console; (3) full-text search API; (4) dashboard service; and (5) graph querying console.

\paragraph{Swarm RESTful API} The SDS provides a RESTful API to manipulate debugging data using the Spring Boot framework\footnote{\url{http://projects.spring.io/spring-boot/}}. Create, retrieve, update, and delete operations are available through HTTP requests and respond with a JSON structure. For example, upon submitting the HTTP request:

\begin{verbatim*}
http://swarmdebugging.org/developers/
search/findByName?name=petrillo
\end{verbatim*}

\noindent the SDS responds with a list of developers whose names are ``petrillo", in JSON format.

\paragraph{SQL Query Console} The SDS provides a console\footnote{\url{http://db.swarmdebugging.org}} to receive SQL queries (SQL) on the debugging data, providing relational aggregations and functions.

\paragraph{Full-text Search Engine} The SDS also provides an ElasticSearch\footnote{\url{https://www.elastic.co/}}, which is a highly scalable open-source full-text search and analytic engine, to store, search, and analyse the debugging data. The SDS instantiates an instance of the ElasticSearch engine and offers a console for executing complex queries on the debugging data.

\begin{figure}[ht]
	\centering
	\includegraphics[width=\linewidth]{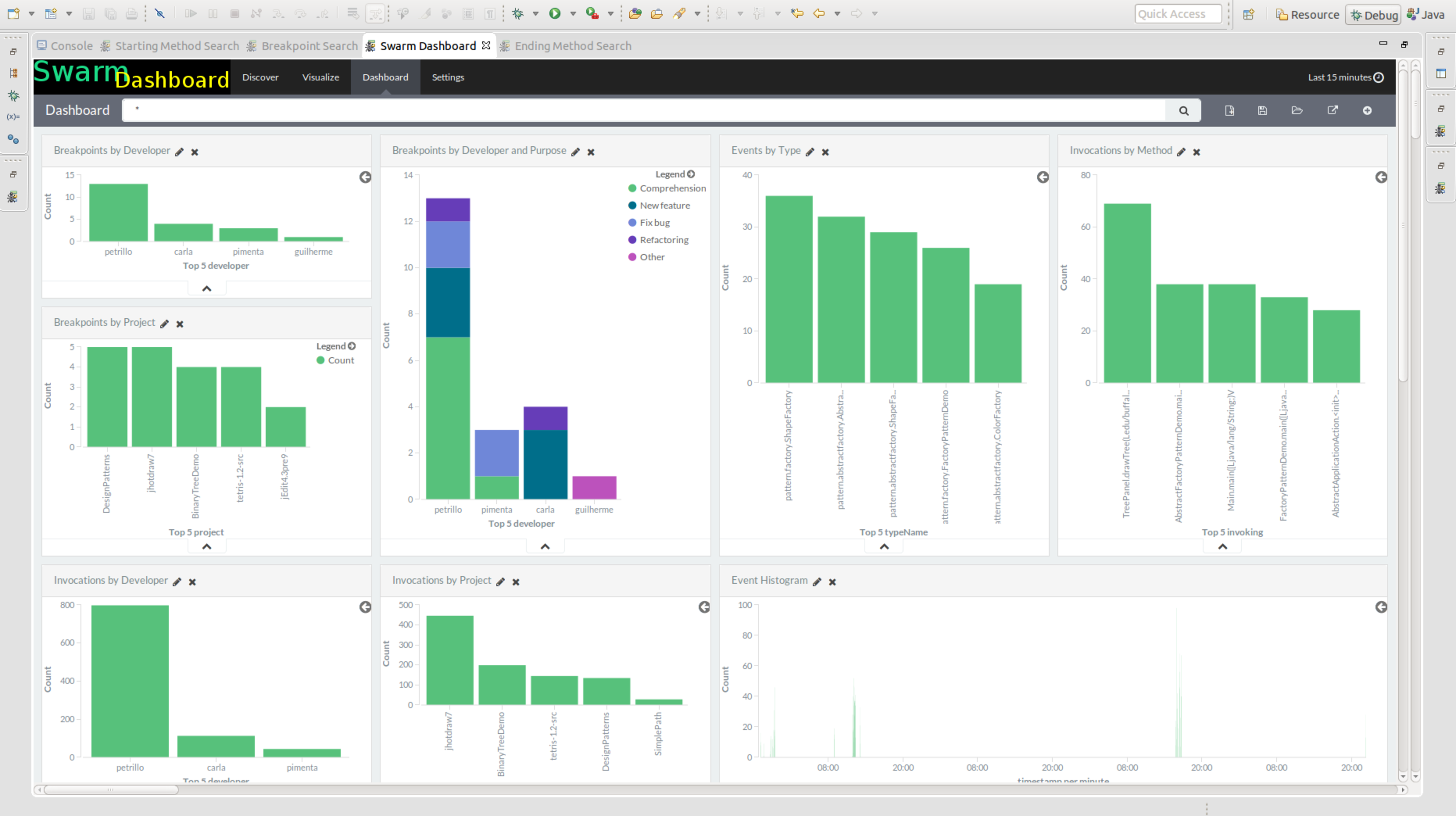}
	\caption{Swarm Debugging Dashboard}
	\label{fig:dashboard}
\end{figure}

\paragraph{Dashboard Service} The ElasticSearch allows the use of the Kibana dashboard. The SDS exposes a Kibana instance on the debugging data. With the dashboard, researchers can build charts describing the data. Figure \ref{fig:dashboard} shows a Swarm Dashboard embedded into Eclipse as a view.

\paragraph{Graph Querying Console} The SDS also persists debugging data in a Neo4J\footnote{\url{http://neo4j.com/}} graph database. Neo4J provides a query language named Cypher, which is a declarative, SQL-inspired language for describing patterns in graphs. It allows researchers to express what they want to select, insert, update, or delete from a graph database without describing precisely how to do it. The SDS exposes the Neo4J Browser and creates an Eclipse view. 

\begin{figure*}
	\centering
	\includegraphics[width=0.8\textwidth]{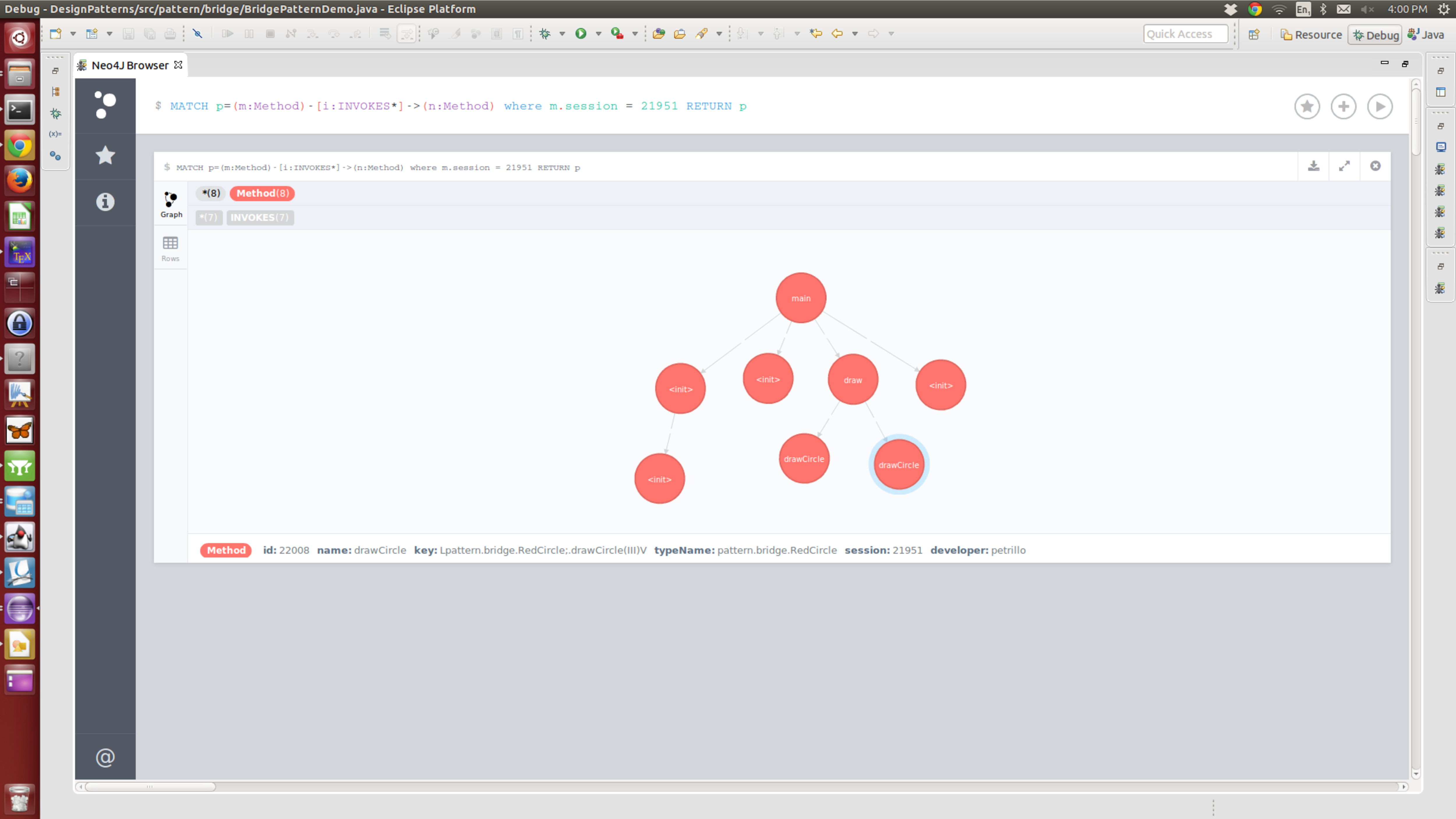}
	\caption{Neo4J Browser - a Cypher query example}
	\label{fig:Neo4JBrowser}
\end{figure*}

Figure \ref{fig:Neo4JBrowser} shows an example of Cypher query and the resulting graph.

\subsection*{Swarm Debugging Tracer}

\begin{figure}[ht]
	\centering
	\includegraphics[width=0.9\linewidth]{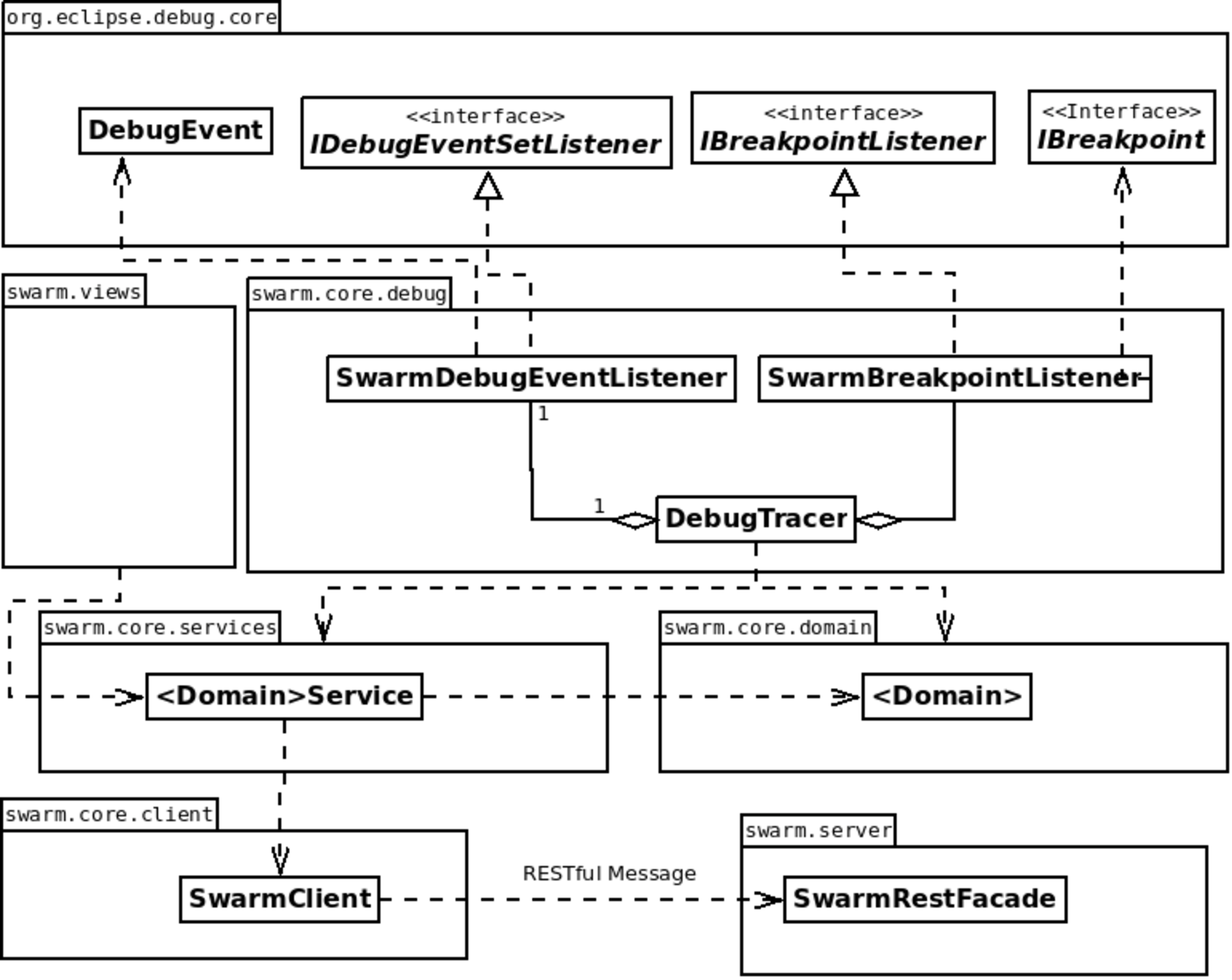}
	\caption{The Swarm Tracer architecture \cite{PetrilloQRS2016}}
	\label{fig:architeture}
\end{figure}

Swarm Debugging Tracer (SDT) is an Eclipse plug-in that listens to debugger events during debugging sessions, extending the Java Platform Debugging Architecture (JDPA). Using the Eclipse JPDA, events are listened by our DebugTracer that implements two listeners:\\ \texttt{IDebugEventSetListener} and \texttt{IBreakpointListener}. Figure \ref{fig:architeture} shows the SDT architecture.

After an authentication process, developers create a debugging session using the Swarm Manager view and toggle breakpoints, trigger stepping events as \textit{Step Into}, \textit{Step Over} or \textit{Step Return}. These events are caught and stack trace items are analyzed by the Tracer, extracting method invocations.

\begin{figure}[ht]
	\centering
	\includegraphics[width=0.7\linewidth]{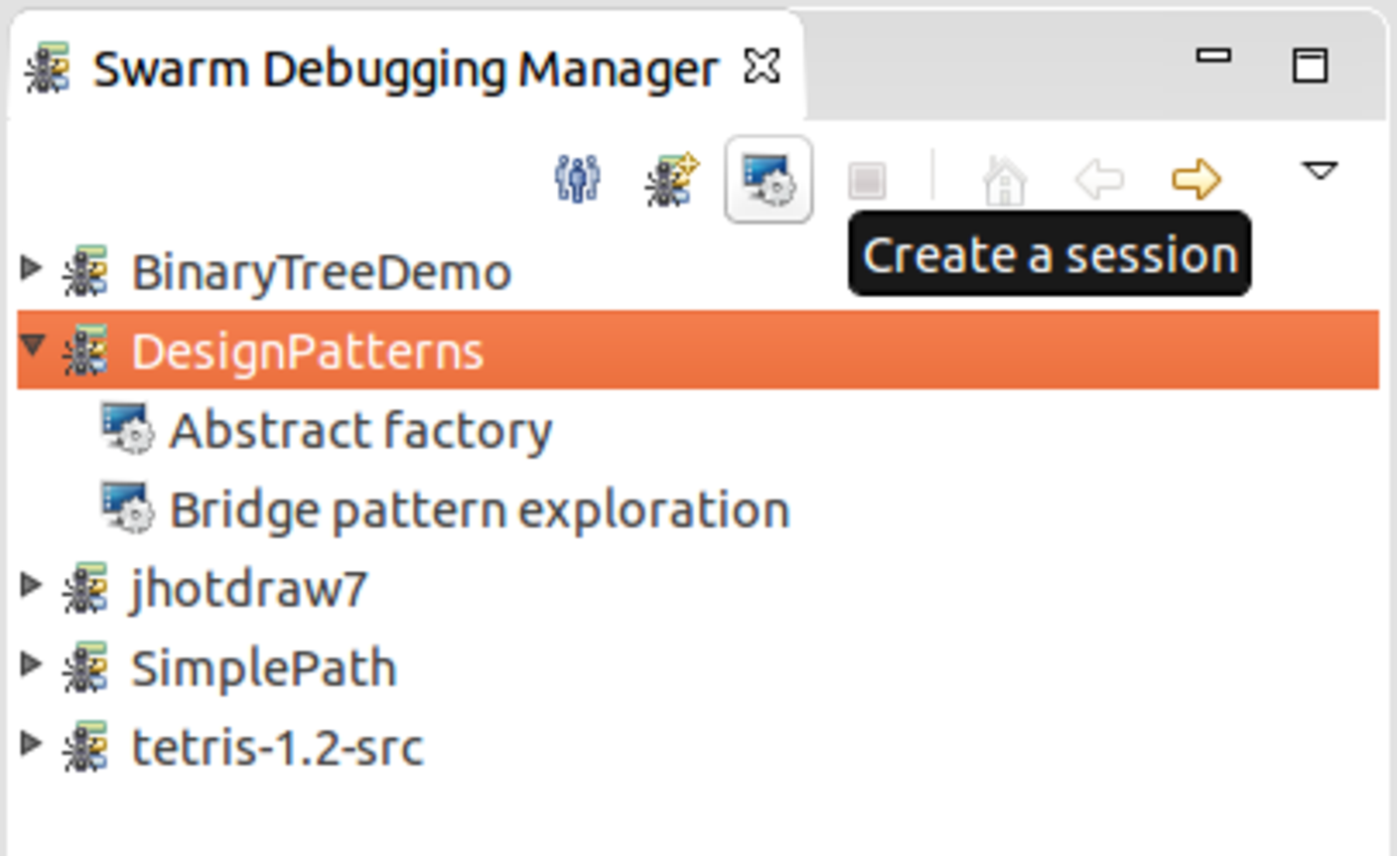}
	\caption{The Swarm Manager view}
	\label{fig:SwarmManager}
\end{figure}

To use the SDT, a developer must open the view ``Swarm Manager" and establish a connection with the Swarm Debugging Services. If the target project is not into the Swarm Manager, she can associate any project in her work-space into Swarm Manager (as shown in Figure \ref{fig:SwarmManager}). This association consists of linking a Swarm Session with a project in the Eclipse workspace. Second, she must create a Swarm session. Once a session is established, she can use any feature of the regular Eclipse debugger, the SDT collects developers' interaction events in the background, with no visible performance decrease.

Typically, the developer will toggle some breakpoints to stop the execution of the program of interest at locations deemed relevant to fix the fault at hand. The SDT collects the data associated to these breakpoints (locations, conditions, and so on). After toggling breakpoints, the developer runs the program in debug mode. The program stops at the first reached breakpoint. Consequently, for each event, such as \textit{Step Into} or \textit{Breakpoint}, the SDT captures the event and related data. It also stores data about methods called, storing invocations entry for each pair invoking/invoked method. Following the foraging approach \cite{Piorkowski2013}, the SDT only collects invoking/invoked methods that were visited by the developer during the debugging session, ignoring other invocations. The debugging activity continues until the program run finishes. The Swarm session is then completed.

To avoid performance and memory issues, the SDT collects and sends the data using a set of specialised \emph{DomainServices} that send RESTful messages to a \emph{SwarmRestFacade}, connecting to the Swarm Debugging Services.

\subsection*{Swarm Debugging Views}

On top of the SDS, the SDI implements and proposes several tools to search and visualise the data collected during debugging sessions. These tools are integrated in the Eclipse IDE, simplifying their usage. They include, but are not limited to the followings.

\begin{figure}
	\centering
	\includegraphics[width=\linewidth]{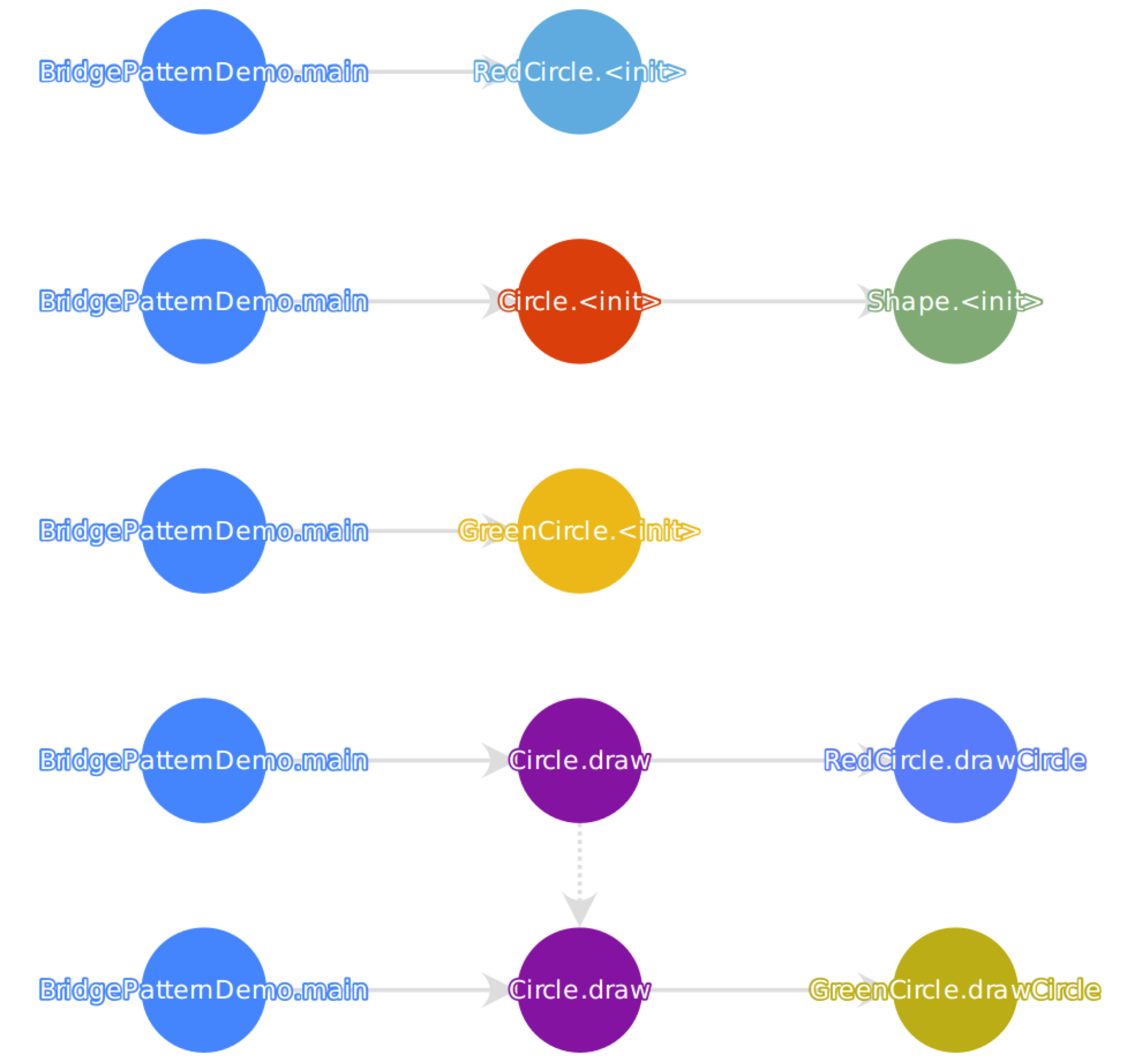}
	\caption{Sequence stack diagram for Bridge design pattern}
	\label{fig:stackdiagram}
\end{figure}

\paragraph{Sequence Stack Diagrams.} Sequence stack diagrams are novel diagrams \cite{Petrillo2015} to represent sequences of method invocations, as shown by Figure \ref{fig:stackdiagram}. They use circles to represent methods and arrows to represent invocations. Each line is a complete stack trace, without returns. The first node is a starting method (non-invoked method) and the last node is an ending method (non-invoking method). If an invocation chain contains a non-starting method, a new line is created and the actual stack is repeated and a dotted arrow is used to represent a return for this node, as illustrated by the method \textit{Circle.draw} in Figure \ref{fig:stackdiagram}. In addition, developers can directly go to a method in the Eclipse Editor by double-clicking over a node in the diagram.

\begin{figure*}
	\centering
	\includegraphics[width=\textwidth]{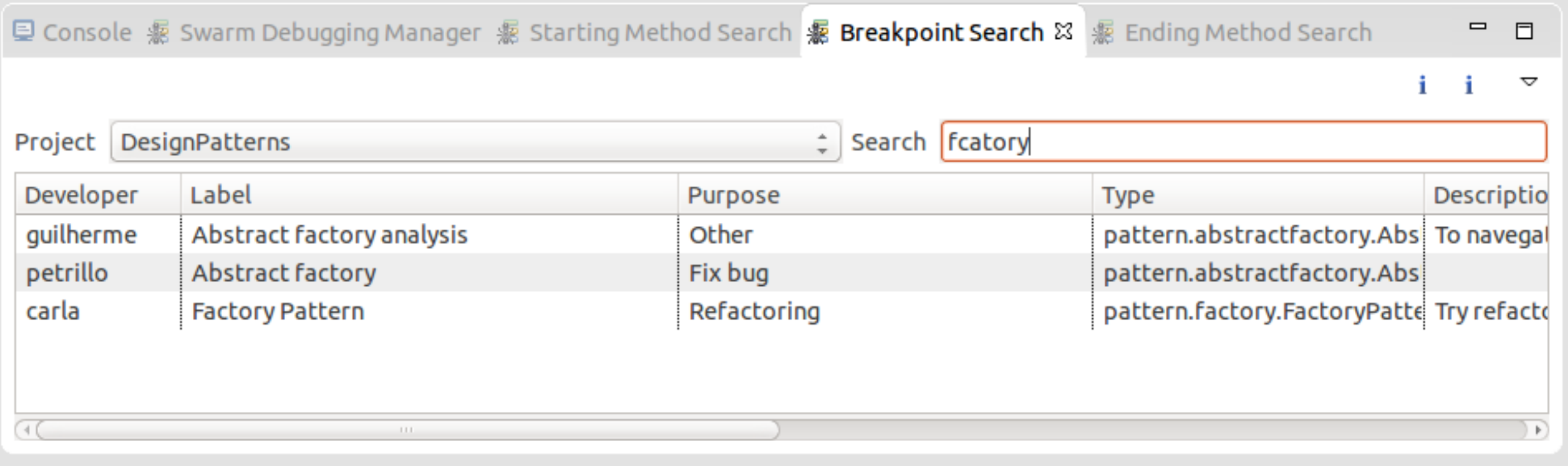}
	\caption{Breakpoint search tool (fuzzy search example)}
	\label{fig:BreakpointSearch}
\end{figure*}

\begin{figure}
	\centering
	\includegraphics[width=\linewidth]{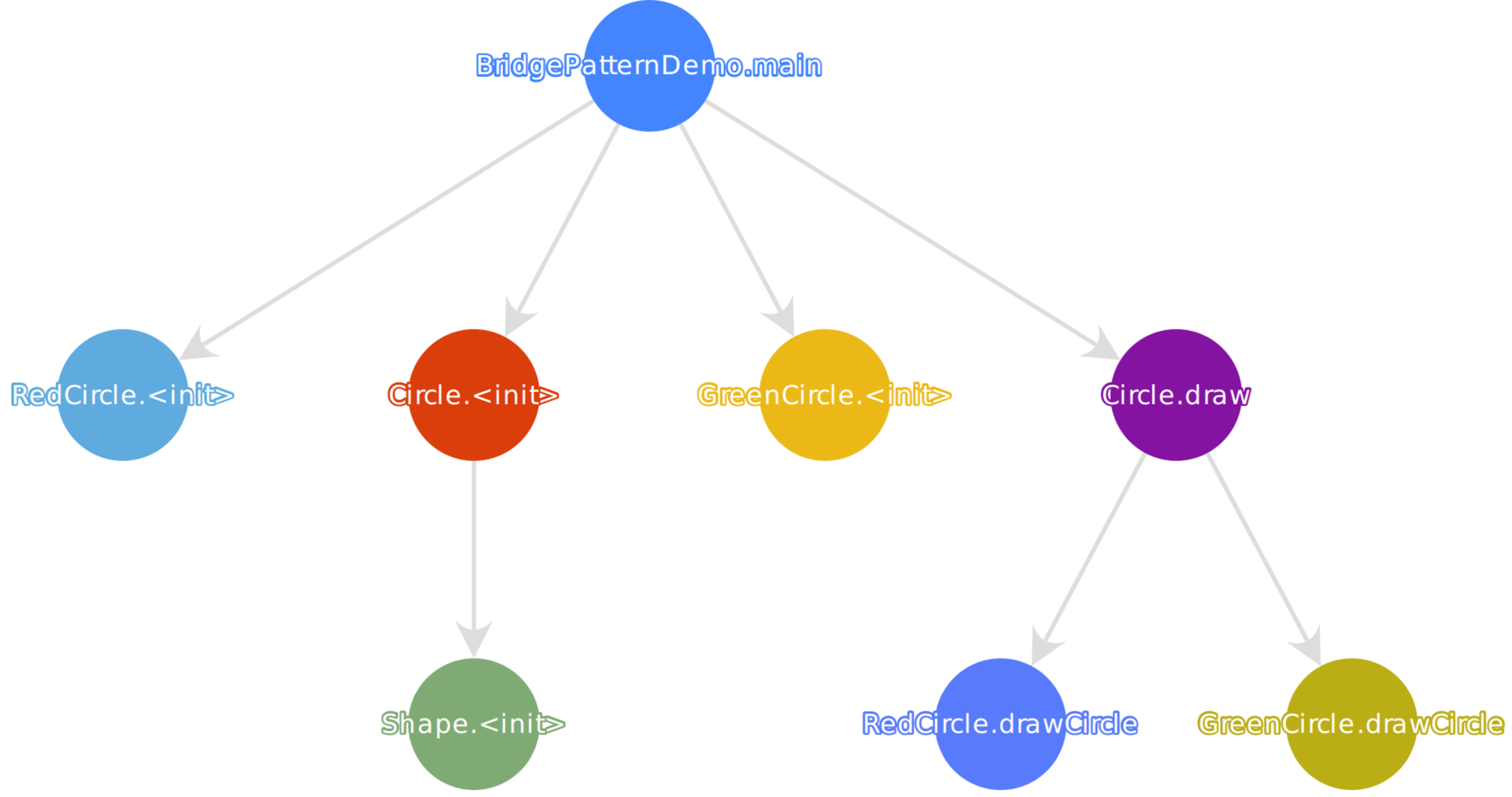}
	\caption{Method call graph for Bridge design pattern \cite{PetrilloQRS2016}}
	\label{fig:callgraph}
\end{figure}

\paragraph{Dynamic Method Call Graphs.} They are direct call graphs \cite{Grove1997}, as shown in Figure \ref{fig:callgraph}, to display the hierarchical relations between invoked methods. They use circles to represent methods and oriented arrows to express invocations. Each session generates a graph and all invocations collected during the session are shown on these graphs. The starting points (non-invoked methods) are allocated on top of a tree and adjacent nodes represent invocations sequences. Researchers can navigate sequences of invocation methods pressing the F9 (forward) and F10 (backward) keys. They can also directly go to a method in the Eclipse Editor by double-clicking on nodes in the graphs.

\subsection*{Breakpoint Search Tool} 

Researchers and developers can use this tool to find suitable breakpoints \cite{Fleming2013} when working with the debugger. For each breakpoint, the SDS captures the type and location in the type where the breakpoint was toggled. Thus, developers can share their breakpoints. The breakpoint search tool allows  \textit{fuzzy}, \textit{match}, and \textit{wildcard} ElasticSearch queries. Results are displayed in the Search View table for easy selection. Developers can also open a type directly in the Eclipse Editor by double-clicking on a selected breakpoint.

Figure \ref{fig:BreakpointSearch} shows an example of breakpoint search, in which the search box contains the misspelled word \textit{fcatory}.

\subsection*{Starting/Ending Method Search Tool}

This tool allows searching for methods that (1) only invoke other methods but that are not explicitly invoked themselves during the debugging session and (2) that are only invoked by others but that do not invoke other methods.

Formally, we define Starting/Ending methods as follows. Given a graph $G=(V,E)$, where $V$ is a set of vertexes $V=\{V_1,V_2,\ldots,V_n\}$ and $E$ is a set of edges $E=\{(V_1,V_2),(V1,V3),\ldots\}$. Then, each edge is formed by a pair: $<V_i,V_j>$, were $V_i$ is the \textit{invoking} method and $V_j$ is the \textit{invoked} method. If $\alpha$ is the subset of all vertexes \textit{invoking} methods and $\beta$ is the subset of all vertexes \textit{invoked} by methods, then the Starting and Ending methods are:

$$StartingPoint = \{V_{SP}~ | ~V_{SP}~ \in \alpha  ~and~ V_{SP} \notin \beta\}$$
$$EndingPoint = \{V_{EP}~ | ~V_{EP}~ \in \beta  ~and~ V_{EP} \notin \alpha\}$$

Locating these methods is important in a debugging session because they are the entries and exits points of a program at runtime.

\subsection*{Summary}

Through the SDI, we provide a technique and model to collect, store and share interactive debugging session data, contextualizing breakpoints and events during these sessions. We created real-time and interactive visualizations using web technologies, providing an automatic memory for developer explorations. Moreover, dividing software exploration by sessions and its call graphs are easy to understand because only intentional visited areas are shown on these graphs, one can go through the execution of a project and see only the important areas that are relevant to developers.

Currently, the Swarm Tracer is implemented in Java, using Eclipse Debug Core services. However, SDI provides a RESTful API that can be accessed independently, and new tracers can be implemented for different IDEs or debuggers.